\documentclass[%
superscriptaddress,
twocolumn,
longbibliography,
amsmath,amssymb,
aps,
prb,
floatfix,
]{revtex4-1}

\usepackage{graphicx}
\usepackage{dcolumn}
\usepackage{bm}
\usepackage{siunitx}
\usepackage{chemformula}
\usepackage{todonotes}
\usepackage{soul}
\usepackage{xcolor}
\usepackage{appendix}
\definecolor{newtext}{RGB}{0, 0, 0}

\bibpunct{[}{]}{,}{n}{}{}

\begin{document}

\preprint{APS/123-QED}

\title{Laser-induced magnetization precession 
\\in individual magnetoelastic domains\\of a multiferroic \ch{CoFeB/BaTiO3} composite
}

\author{L.\,A.\,Shelukhin}\email{shelukhin@mail.ioffe.ru}
 \affiliation{Ioffe Institute, 194021 St. Petersburg, Russia}
\author{N.\,A.\,Pertsev}
 \affiliation{Ioffe Institute, 194021 St. Petersburg, Russia}
 \author{A.\,V.\,Scherbakov}
 \affiliation{Ioffe Institute, 194021 St. Petersburg, Russia}  \affiliation{Experimental Physics II, Technical University Dortmund, D-44227 Dortmund, Germany}
 \author{D.\,L.\,Kazenwadel}
 \affiliation{University of Konstanz, D-78457 Konstanz, Germany}
  \author{D.\,A.\,Kirilenko}
 \affiliation{Ioffe Institute, 194021 St. Petersburg, Russia}
 \author{S.\,J.\,H\"{a}m\"{a}l\"{a}inen}
\affiliation{NanoSpin, Department of Applied Physics, Aalto University School of Science, P.O. Box 15100, FI-00076 Aalto, Finland}
 \author{S.\,van\,Dijken}
\affiliation{NanoSpin, Department of Applied Physics, Aalto University School of Science, P.O. Box 15100, FI-00076 Aalto, Finland}
 \author{A.\,M.\,Kalashnikova}
 \affiliation{Ioffe Institute, 194021 St. Petersburg, Russia}

\date{\today}

\begin{abstract}
Using a magneto-optical pump-probe technique with micrometer spatial resolution we show that magnetization precession can be launched in individual magnetic domains imprinted in a Co$_{40}$Fe$_{40}$B$_{20}$ (CoFeB) layer by elastic coupling to ferroelectric domains in a \ch{BaTiO3} substrate.
The dependence of the precession parameters on external magnetic field strength and orientation reveal that by laser-induced ultrafast partial quenching of the magnetoelastic coupling parameter of CoFeB by \textcolor{newtext}{$\approx$~27\,\% along with 10\,\%} ultrafast demagnetization trigger the magnetization precession.
The relation between the laser-induced reduction of the magnetoelastic coupling and the demagnetization is
approximated by the $n(n+1)/2$-law with \textcolor{newtext}{$n\approx$~2}.
This correspondence confirms the thermal origin of the laser-induced anisotropy change.
\textcolor{newtext}{Based on the analysis and modeling of the excited precession we find signatures of laser-induced precessional switching, which occurs when the magnetic field is applied along the hard magnetization axis and its value is close to the effective magnetoelastic anisotropy field.}
The precession excitation process in an individual magnetoelastic domain is found to be unaffected by neighboring domains. This makes laser-induced changes of magnetoelastic anisotropy a promising tool for driving magnetization dynamics \textcolor{newtext}{and switching} in composite multiferroics with spatial selectivity.
\end{abstract}

\pacs{Valid PACS appear here}
\maketitle

\section{Introduction}

Magnetoelectric multiferroics offer a possibility to control magnetization by an electric field \cite{Smolenskii:1982,Fiebig:2005,Hu:2015}, which is in high demand for beyond CMOS technologies \cite{Manipatruni-Nature2019} including sensors, energy harvesters, memories and logic devices \cite{Chu-JPD2018}.
However, the number of single-phase multiferroics is limited \cite{Hill-JPCB2000}, with many of the materials exhibiting multiferroic properties below room temperature or providing insufficient coupling between the order parameters.
On the other hand, composite structures consisting of ferromagnetic (FM) and ferroelectric (FE) materials coupled via strain represent a promising alternative for achieving indirect magnetoelectric coupling at room temperature of sufficient strength for future applications \cite{Vaz:2012,Wang-NPGAsia2010,carman-MRS2018,Chu-JPD2018,Nan:JAP2008}.

In strain-coupled composite multiferroics, the magnetic anisotropy of the FM component is altered by a combination of interfacial strain transfer from the FE one and inverse magnetostriction.
For properly selected FM and FE materials, heterostructure geometries, and optimized interfaces, strong magnetization responses to external electric fields can be obtained.
In particular, full imprinting of ferroelastic domain patterns from FE substrates into FM overlayers with in-plane and perpendicular magnetizations and their subsequent manipulation by electric fields have been demonstrated experimentally \cite{Taniyama-JAP2007,Lahtinen-AdvMat2013,shirahata-NPGAsia2015,lahtinen-SciRep2012,buzzi-PRL2013,franke-PRX2015,LoContePRMater:2018}. FM-FE composites enable electric-field-induced magnetization switching leading to changes in magnetoresistance \cite{Cavaco2007,Pertsev-APL2009,Pertsev-Nanotech2010,LiuAPL:2011,Chen-NComm2019}, electrical tuning of ferromagnetic resonance and spin-wave spectra \cite{liu-AdvMat2013,LouAdMat:2009,Brandl:2014,Azovtsev-PRAppl2018}, active filtering and routing of propagating spin waves \cite{HamalainenNature:2018,sadovnikov-PRL2018}, and electrical switching between superconducting and normal states \cite{Savostin-Nanoscale2020}.

Control over the order parameters of composite multiferroic structures by femtosecond laser pulses would widen their application perspectives and elevate their switching speed.
There are several pathways along which an ultrafast optical stimulus may alter the state of a composite multiferroic.
One relies on ultrafast direct optical control of magnetization \cite{Kirilyuk-RMP2010} and, via FM-FE coupling, of ferroelectric polarization \cite{Sheu:2014,Jia:2016}.
Alternatively, ultrafast optically-driven changes of the FE state could lead to high-amplitude dynamical strain modulations \cite{Lejman-NatureComm2014} and, thus, alter the magnetization state of a magnetostrictive layer.
However, ultrafast optical control of FE polarization remains challenging \cite{KIMEL2020}.
Optically-induced strain as a tool to modify the magnetic anisotropy in FM-photostrictive composites is considered in \cite{Liu-ACSNano2012}, but results reported thus far do not extend to ultrafast timescales.

In this Article, we examine an alternative approach to ultrafast optical control of a composite multiferroic, and study the feasibility of ultrafast laser-induced changes of strain-mediated magnetoelectric coupling in a \ch{CoFeB/BaTiO3} heterostructure.
Using a femtosecond magneto-optical pump-probe technique with micrometer spatial resolution, we excite magnetization precession in individual magnetic domains imprinted in the amorphous CoFeB layer by mechanical coupling to ferroelastic domains in BaTiO$_3$. 
We reveal that the precession is triggered by an abrupt decrease of the CoFeB magnetoelastic coupling parameter and magnetization by \textcolor{newtext}{ about $27$\,\% and $10$\,\%}, respectively, for an incident laser pulse fluence of 10\,mJ/cm$^2$. 
This ratio satisfies the $n(n+1)/2$-law with $n\approx2$ that describes the temperature-induced variations of a single-ion uniaxial magnetic anisotropy, confirming the thermal origin of the observed changes. \textcolor{newtext}{When the magnetic field is applied along the hard anisotropy axis and is comparable to the effective magnetoelastic anisotropy field, we find signatures of precessional switching of magnetization. The switching manifests itself in the suppression of the detected pump-probe signal.}

\textcolor{newtext}{The Article is organized as follows.
In Sec.\,\ref{sec:experimental} we describe the studied CoFeB/BaTiO$_3$ heterostructure and the experimental procedures.
In Sec.\,\ref{Sec:results} we present experimental results on laser-induced magnetization precession in individual domains of CoFeB/BaTiO$_3$.
In Sec.\,\ref{Sec:discussion} we present an analysis of the laser-induced precession based on a thermodynamic approach and the Landau-Lisfitz-Gilbert equation.
This is followed by the Conclusions Section wherein we also outline possible applications of laser-induced control of magnetoelastic anisotropy in spintronics.}

\section{Experimental}\label{sec:experimental}

\subsection{CoFeB/BaTiO$_3$ heterostructure}

\begin{figure}
\includegraphics[width=0.5\textwidth]{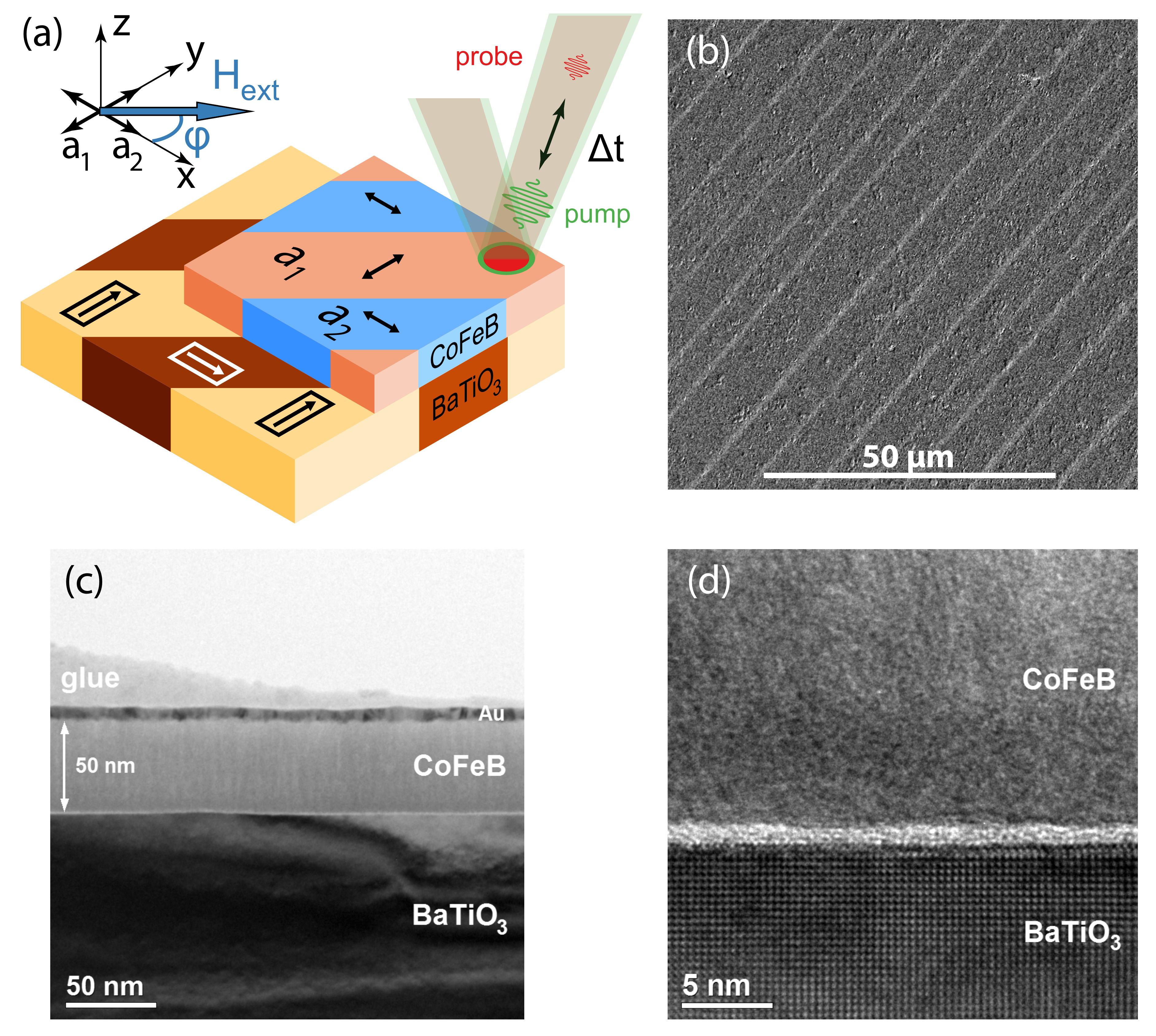}
\caption{(Color online) (a) Schematics of the \ch{CoFeB/BaTiO3} structure and the pump-probe experiment.
Arrows in rectangles and double-headed arrows indicate the spontaneous polarization in the ferroelastic domains of BaTiO$_3$ and the strain-induced magnetic anisotropy axes of the a$_1$ and a$_2$ domains in the CoFeB layer, respectively. (b) Magneto-optical image of the sample obtained using the longitudinal Kerr microscopy. (c) Bright-field and (d) high-resolution TEM images of the CoFeB/BaTiO$_3$ heterostructure.
}
\label{Fig:Sample}
\end{figure}

The heterostructure under study consists of a 50-nm-thick layer of a FM CoFeB amorphous alloy on a \SI{500}{\micro\meter}-thick single-crystalline \ch{BaTiO3} (001) substrate.
The CoFeB layer was grown on {BaTiO$_3$} by magnetron sputtering at $T_{g}=$~573~K and capped with a 6-nm-thick Au layer.
\textcolor{newtext}{The amorphous nature of the CoFeB film was verified by transmission electron microscopy.
Figure\,\ref{Fig:Sample}~(c,d) shows TEM images of the CoFeB/BaTiO$_3$ heterostructure.
Cross-section TEM specimen were prepared by mechanical polishing and subsequent Ar$^+$ ion milling at 3~keV.
A Jeol JEM-2100F microscope operated at 200~kV was used to acquire images in conventional bright-field and high-resolution modes.
TEM images indicate that the CoFeB film is amorphous with a short-range order corresponding to a lengthscale of $\sim$1\,nm.}

At room temperature (RT), the {BaTiO$_3$} substrate is split into \ang{90} ferroelectric-ferroelastic stripe domains with in-plane spontaneous polarization aligned along the long side of the tetragonal unit cell \cite{Lahtinen-AdvMat2013}.
Owing to the strain transfer from BaTiO$_3$ to CoFeB and inverse magnetostriction, an uniaxial magnetoelastic anisotropy is locally induced in the CoFeB film.
Since the magnetostriction parameter $\lambda$ of CoFeB is positive, the magnetic anisotropy easy axes are oriented parallel to the polarizations of the underlying FE domains \cite{lahtinen-SciRep2012,Lahtinen-AdvMat2013} [Fig.~\ref{Fig:Sample}~(a)].
The strain-induced magnetoelastic anisotropy dominates over other anisotropy contributions and, consequently, the stripe domains in CoFeB and \ch{BaTiO3} fully correlate.

\textcolor{newtext}{The magnetoelastic domain pattern in CoFeB was imaged by magneto-optical Kerr microscopy (see \cite{Lahtinen-AdvMat2013} for details).
As can be seen from the image in Fig.\,\ref{Fig:Sample}~(b), the two types of domains have different widths.
The width of the wider domain varies around \SI{12}{\micro\meter}, while the width of the narrow domain is about \SI{3}{\micro\meter}.
The two domain types are labeled a$_1$ and a$_2$ [see Fig.\,\ref{Fig:Sample}~(a)].} 

\subsection{Magneto-optical pump-probe setup}

Laser-induced magnetization dynamics in the distinct magnetic domains of the CoFeB/\ch{BaTiO3} composite was studied using a femtosecond two-color magneto-optical pump-probe setup.
In the experiments, the pump and probe pulse duration was 170\,fs, the central wavelengths of the pump and probe pulses were 515\,nm and 1030\,nm, the pump pulse fluence was 10\,mJ/cm$^2$, and the probe pulse fluence was $\sim$10 times lower.
Both pump and probe pulses were focused on the CoFeB layer 
into a spot with a diameter below \SI{5}{\micro\meter} using a 15x reflective microscope objective.
The laboratory frame was chosen such that the $x$, $y$, and $z$ axes are directed along the easy axes of the a$_2$, a$_1$ domains and the sample normal, respectively [Fig.\,\ref{Fig:Sample}~(a)].
An external DC magnetic field $\mathbf{H}_\mathrm{ext}$ of strength 0--120\,mT was applied in the sample plane at an angle $\varphi$ to the $x$ axis.
Measurements of the laser-induced magnetization dynamics at $\varphi=0,\pm\ang{45}$ in the a$_1$ and a$_2$ domains were performed by displacing the sample laterally with \SI{0.05}{\micro\meter} precision.
The pump-induced changes of magnetization were traced by recording the magneto-optical Kerr rotation $\Delta\theta$ of the probe polarization plane as a function of time delay $\Delta t$ between the pump and probe pulses.
The incidence angle of the probe pulses was $\zeta=$~\ang{17}, and the measured Kerr rotation $\Delta\theta(\Delta t)$ was proportional mostly to the pump-induced changes of the out-of-plane component $M_{z}$ of magnetization.
All measurements were performed at RT.

\textcolor{newtext}{Additional static magneto-optical characterization and pump-probe studies of laser-induced demagnetization were performed with larger laser spots.
Since the easy magnetization axes lie in the sample plane, the static magneto-optical and the ultrafast demagnetization measurements were conducted in the longitudinal magneto-optical Kerr effect (MOKE) geometry with a probe incidence angle $\zeta=$~\ang{45}.
The pump was incident along the sample normal.
Long-focal distance lenses were used to focus pump and probe pulses on the sample surface into spots with diameters of \SI{60}{\micro\meter} and \SI{30}{\micro\meter}, respectively.
Since the spot sizes in these experiments exceed the widths of both the a$_1$ and a$_2$ domains, the external magnetic field $H_\mathrm{ext}=150$\,mT was applied along the a$_1$ anisotropy easy axis to fully saturate the magnetization of the sample into a single-domain state.
In this case, the MOKE signal is proportional to the saturation magnetization $M_S$.
The incident probe pulses were $p$-polarized to maximize the longitudinal MOKE response.
Static MOKE measurements without pump pulses reveal a static Kerr rotation $\theta^\mathrm{L}_S=2.5$\,mdeg and a Kerr ellipticity $\epsilon_S=20$\,mdeg in saturation. 
To study ultrafast demagnetization, laser-induced dynamics of the probe ellipticity $\Delta\epsilon(\Delta t)\sim\Delta M_S(\Delta t)$ was measured upon excitation of the sample by pump pulses with a fluence $F=10$\,mJ/cm$^{2}$.
The degree of ultrafast demagnetization was calculated as $\Delta M_S/M_S(\Delta t)=0.5[\Delta\epsilon(\Delta t,\mathrm{+H_{ext}})-\Delta\epsilon(\Delta t,\mathrm{-H_{ext}})]/\epsilon_S$.}

\textcolor{newtext}{The longitudinal Kerr rotation $\theta^\mathrm{L}_S$ and Kerr ellipticity $\epsilon_S$  were also used to estimate the polar Kerr rotation $\theta^\mathrm{P}_S\approx15$\,mdeg corresponding to magnetization saturation along the perpendicular direction (see App.\,\ref{App:MOKE} for details).
$\theta^\mathrm{P}_S$ was used to normalize the dynamical Kerr rotation $\theta(\Delta t)$ obtained in the pump-probe experiments and to obtain the magnitude of the laser-induced change of magnetization along the $z-$axis $\Delta M_z(\Delta t)/M_s=\Delta\theta(\Delta t)/\theta^\mathrm{P}_S$.
We note that the estimated value of $\theta^\mathrm{P}_S$ contains an uncertainty [see App.\,\ref{App:MOKE}].}

\section{Results}\label{Sec:results}

\begin{figure}
\includegraphics[width=0.49\textwidth]{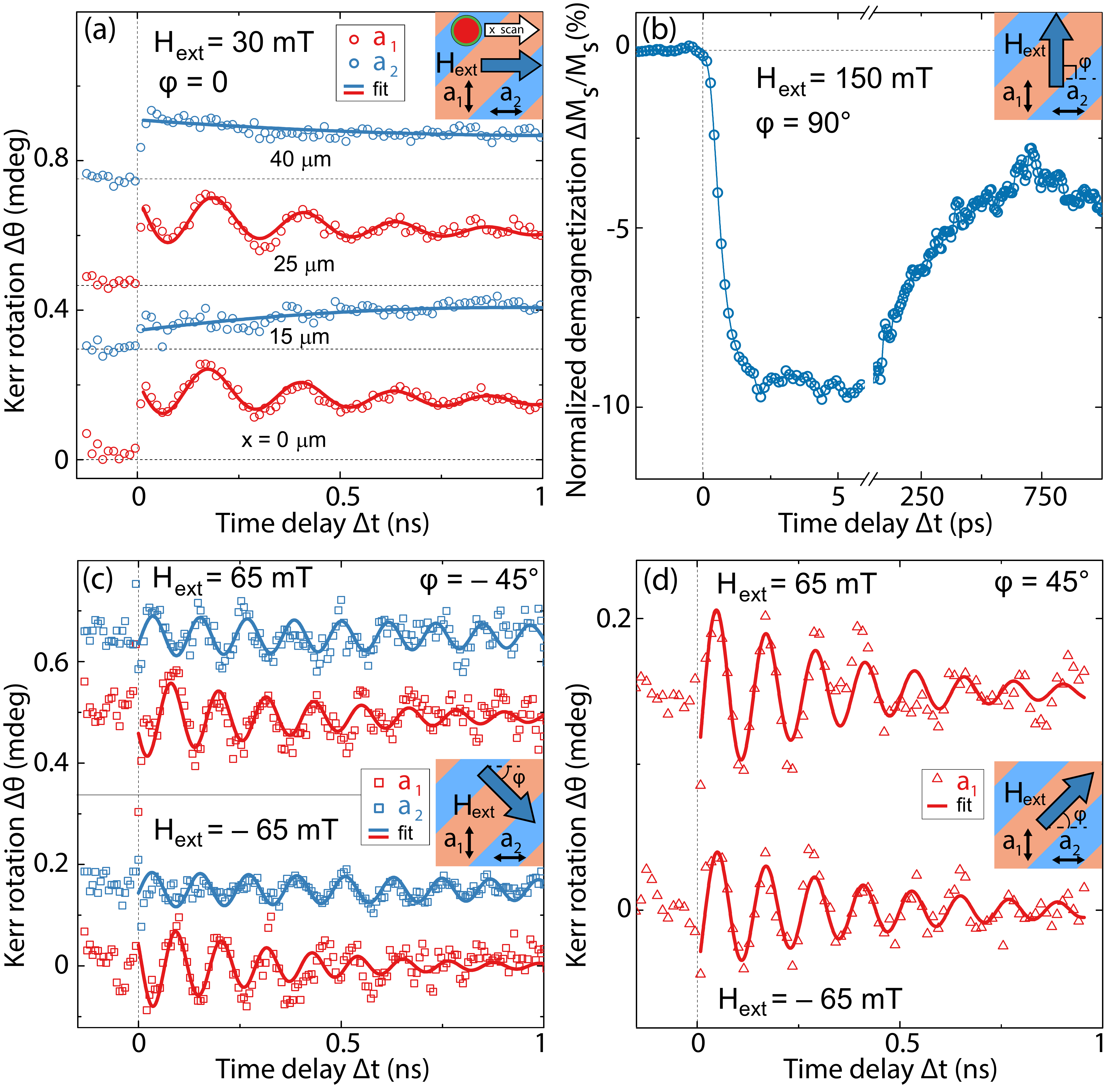}
\caption{(Color online) (a,~c,~d) Laser-induced Kerr rotation $\Delta\theta$ of the probe pulse polarization as a function of time delay $\Delta t$ measured in a$_1$ (red symbols) or a$_2$ (blue symbols) domains.
The external magnetic field $\mathbf{H}_\mathrm{ext}$ is applied at (a) $\varphi=$~0, (c) $\varphi=$~\ang{-45}, and (d) $\varphi=$~\ang{45}.
In (c,d) the signals measured at positive (upper curves) and negative (lower curves) $H_\mathrm{ext}$ are shown.
Lines are fits to the measurement data (see text).
(b) Ultrafast demagnetization measured at $H_\mathrm{ext}=$~150\,mT.
}
\label{Fig:All_Measurements}
\end{figure}

Figure\,\ref{Fig:All_Measurements}~(a) shows the laser-induced probe polarization rotation $\Delta\theta(\Delta t)$ measured for different positions $x$ of laser spots on the sample surface.
\textcolor{newtext}{We note that translation along the $x-$axis corresponds to a translation by $x\sqrt{2}/2$ along the direction normal to the domain wall.}
The external magnetic field $\mathbf{H}_\mathrm{ext}$ is parallel (perpendicular) to the easy axis of the a$_{2}$ (a$_{1}$) domain ($\varphi=0$).
At $H_\mathrm{ext}=30$\,mT one clearly distinguishes two types of dynamic signal $\Delta\theta(\Delta t)$ depending on $x$.
At $x=$\,0 and \SI{25}{\micro\meter} clear oscillations of $\Delta\theta$ are observed, while at $x=$\,\SI{15} and \SI{40}{\micro\meter} only a slowly varying change of $\Delta\theta$ is seen.
Detailed studies of $\Delta\theta(\Delta t)$ at $x=$\,0\,for different magnetic field strengths reveal that the oscillatory signal is present only in the range of 0--45\,mT.
The amplitude and the frequency of the oscillations as a function of the applied magnetic field were obtained by fitting the signal at $x=$\,0 to the function $\Delta\theta(\Delta t)=\Delta\theta_0\exp(-t/\tau_\mathrm{d})\sin(2\pi ft+\xi_0)+P_2(t)$, where $\Delta\theta_\mathrm{0}$, $f$, $\xi_0$, $\tau_\mathrm{d}$ are the oscillation amplitude, frequency, initial phase, and decay time, respectively.
The second-order polynomial function $P_2(t)$ accounts for a slowly varying background of nonmagnetic nature.

\begin{figure}
\includegraphics[width=0.5\textwidth]{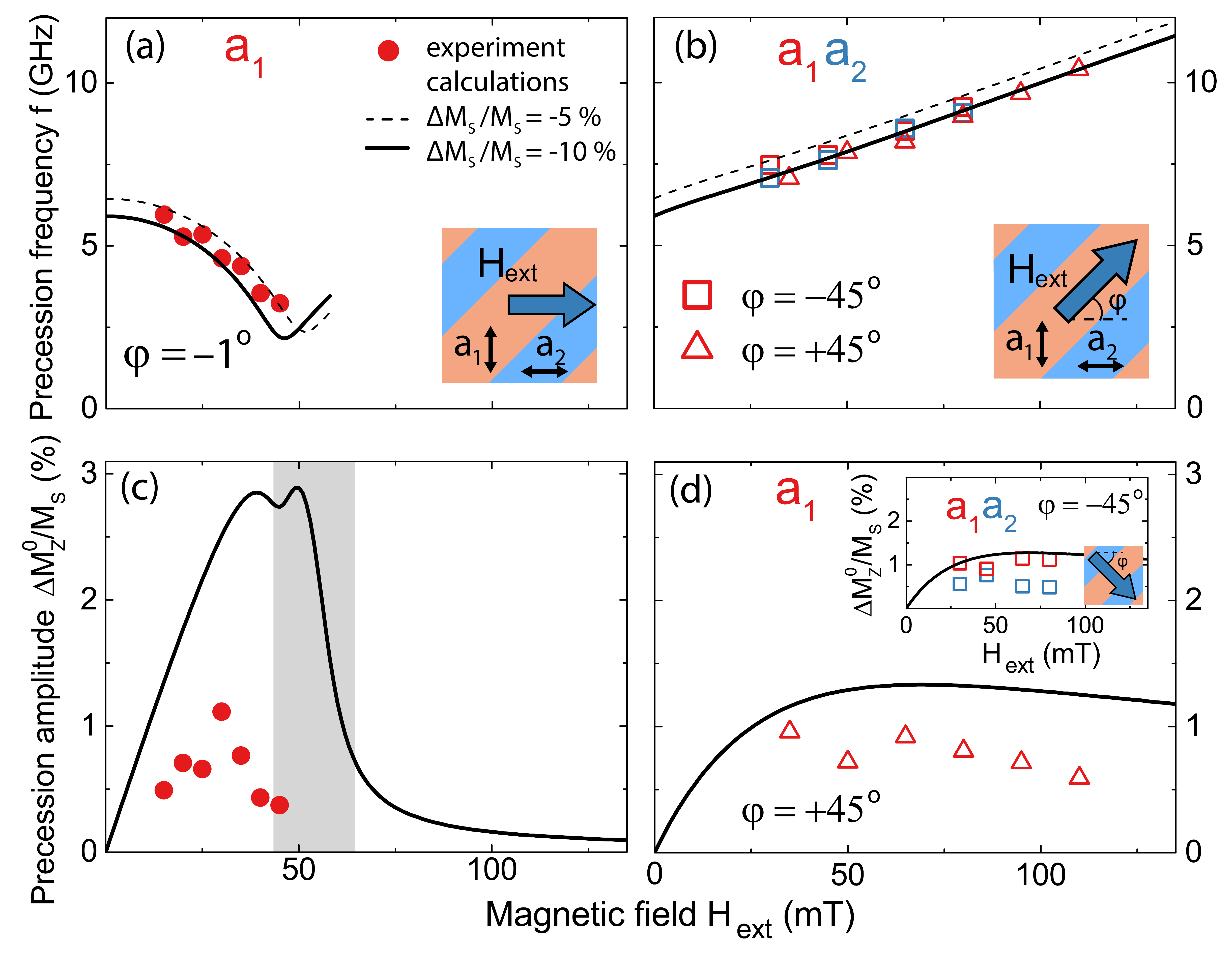}
\caption{(Color online) Experimental magnetic field dependencies of the precession frequency $f$ (a,~b) and  the normalized amplitude $\Delta M^0_z/M_S$ (c,~d) in the a$_{1}$ (red symbols) and a$_{2}$ (blue symbols) domains at (a,~c) $\varphi$~=~\ang{-1} , (b,~d) $\varphi=$~\ang{-45} (open squares), and  $\varphi=$~\ang{45} (triangles). Solid and dashed lines are the calculated dependencies for the a$_1$ domain for $\Delta M_S/M_S=10\%$ and $\Delta M_S/M_S=5\%$, respectively.
}
\label{Fig:Amps_Freqs}
\end{figure}

The field dependence of the frequency $f$ [Fig.\,\ref{Fig:Amps_Freqs}~(a)] resembles that of a magnetization precession in a field applied perpendicularly to the easy axis.
Therefore, the signal at $x=$\,0 can be confidently ascribed to the laser-induced precession of magnetization in the a$_1$ domain.
The periodicity and the width of the areas in which laser-induced precession is detected (at $x=0,\,25$\,$\mathrm{\mu}$m) and not detected (at $x=15,\,40$\,$\mathrm{\mu}$m) correspond to the length scale of the magnetic domain pattern in the sample.
\textcolor{newtext}{The field dependence of the change in the out-of-plane magnetization component $\Delta M_z^0/M_S=\Delta\theta^0/\theta^P_S$ is shown in Fig.\,\ref{Fig:Amps_Freqs}~(c).}

If the external field $\mathbf{H}_\mathrm{ext}$ is applied at an angle $\varphi=$\ang{-45}, i.e., if it makes an angle of $\pm$\ang{45} to the easy axes of the a$_1$ and a$_2$ domains, a laser-induced precession is observed in both stripe domains in a wider field range. 
Fig.\,\ref{Fig:All_Measurements}~(c) shows $\Delta\theta(\Delta t)$ measured in the a$_{1}$ (red dots) and the a$_{2}$ (blue dots) domains at $H_\mathrm{ext}=\pm$65\,mT.
As one can see, upon the transition from the a$_1$ to the a$_2$ domain the initial phase $\xi_0$ of the magnetization precession changes by \ang{180}.
The change of the magnetic field sign, in turn, does not affect the character of the excited precession. 
The dependence of the precession frequency on field strength [Fig.\,\ref{Fig:Amps_Freqs}~(b)] is typical for a geometry wherein the field is applied at \ang{45} with respect to the uniaxial magnetic anisotropy axis and, bacause of symmetry, is the same for both domains. 
The precession amplitude shows small variations with applied field strength and is somewhat higher in the wider a$_1$ domain [inset Fig.\,\ref{Fig:Amps_Freqs}~(d)]. 
When the external field is applied at $\varphi$~=~\ang{45}, 
a laser-induced precession is detected in the a$_{1}$ domains only [Fig.\,\ref{Fig:All_Measurements}~(d)]. 
The field dependencies of the precession frequency and amplitude are similar to those found in the a$_1$ domains at $\varphi=$~\ang{-45} [Fig.\,\ref{Fig:Amps_Freqs}~(b,d)].

The magnetization precession in the individual domains excited at $\varphi=$~0,\ang{-45} agrees with the general scenario of laser-induced changes of magnetic anisotropy \cite{Bigot-ChemPhys2005,CarpenePRB:2010}.
As discussed in detail in e.g. \cite{CarpeneJAP:2010,Shelukhin:PRB2018}, a fast change of the effective anisotropy field produces a magnetization precession.
The efficiency of excitation is highly sensitive to the alignment of equilibrium magnetization with respect to the easy anisotropy axis.
In our experiments, the magnetization precession in a particular domain is not excited when the external field is aligned along the easy anisotropy axis of the domain [Fig.\ref{Fig:All_Measurements}~(a)].
When the field is aligned perpendicularly to the easy anisotropy axis of the domain, the precession is excited only if the field strength is below the anisotropy field.
The latter is easily identified in our experiments as the field at which the frequency of the precession is minimal [Fig.\,\ref{Fig:Amps_Freqs}~(a)].
Finally, when the magnetic field is at an intermediate angle to the easy anisotropy axis, e.g. at $\pm$\ang{45}, the magnetization precession is excited in a wider field range [Fig.\ref{Fig:Amps_Freqs}].

\textcolor{newtext}{Below we use a macrospin model to analyse the process of excitation and the following decay of the magnetization precession in order to identify the underlying excitation mechanism.
In the analysis we mostly focus on the a$_1$ domain and neglect effects on the laser-induced dynamics imposed by dipolar coupling to neighboring domains. 
This is justified by the experimental results showing that the frequencies of the excited precession in a$_1$ and a$_2$ domains at $\varphi=\pm$\ang{45} are equal.
The difference in the precession amplitudes detected in the a$_{1}$ and a$_{2}$ domains at $\varphi=$~\ang{-45} [Figs.\,\ref{Fig:All_Measurements}~(c) and \ref{Fig:Amps_Freqs}~(d)] could originate from the fact that the width of the a$_{2}$ domain is somewhat smaller than the probe spot size.
As a result, the probe measures the magnetization dynamics of the a$_2$ domain and a part of the neighbouring a$_{1}$ domains, where the precession phase $\xi_0$ is opposite. 
This reduces the overall detected transient Kerr rotation.
The effect is even more pronounced at $\varphi=$~\ang{+45}, where no precession signal from the a$_2$ domain is detected.
In this case, the field is applied along the stripe domains, which initializes head-to-head and tail-to-tail domain walls having a width of about \SI{1.5}{\micro\meter} \cite{HamalainenNature:2018}.
As it is comparable to the width of the a$_2$ domain, the magnetization in the narrow a$_2$ domain is non-uniform, which suppresses the measurement signal.
For the wider a$_1$ domains or field orientations that initialize narrow head-to-tail domain walls, a similar reduction of the precession signal does not occur.}

\section{Discussion}\label{Sec:discussion}

\textcolor{newtext}{\subsection{Magnetic anisotropy of the CoFeB/BaTiO$_3$ heterostructure}}

In contrast to metallic films with a pronounced magnetocrystalline anisotropy considered in \cite{Bigot-ChemPhys2005,CarpenePRB:2010} and following works, the magnetic anisotropy of amorphous CoFeB film is fully dominated by an uniaxial magnetoelastic anisotropy.
Hence the magnetization-dependent part of the free energy density $F$ in an individual domain contains only the Zeeman, magnetoelastic, and shape anisotropy terms and can be approximated by the relation \cite{Pertsev-PRB2008}:
\begin{eqnarray}
    F&=&- \mu_0\mathbf{M}_S\cdot\mathbf{H}_\mathrm{ext}+B_1\left(u_{xx}m_x^2+u_{yy}m_y^2\right)\nonumber\\
     &+&\left[\frac{1}{2}\mu_{0}M_{S}^2-\frac{c_{12}}{c_{11}}B_1(u_{xx}+u_{yy})\right]m_{z}^2,\label{eq:energy}
\end{eqnarray}
where $m_i=M_i/M_S$, $u_{xx}$, $u_{yy}$ are the substrate-induced in-plane strains in the CoFeB film, $B_1=-1.5\lambda(c_{11}-c_{12})$ is the magnetoelastic coupling parameter, and $c_{11}=2.8\cdot10^{11}$\,N/m$^2$, $c_{12}=1.4\cdot10^{11}$\,N/m$^2$ are the elastic stiffnesses of CoFeB at constant magnetization $\mathbf{M}_S$ taken to be those of Fe$_{60}$Co$_{40}$ \cite{CoFe-elastic}.
The misfit strains $u_{xx}$ and $u_{yy}$ are expected to be fully relaxed at the growth temperature $T_g=573$\,K, which is well above the Curie temperature $T_C=393$\,K of BaTiO$_3$.
On cooling from $T_g$, nonzero strains appear in the film owing to the difference in the thermal expansion coefficients of the paraelectric BaTiO$_3$ $\alpha_0 = 10\cdot10^6$~K$^{-1}$ \cite{Touloukian} and CoFeB $\alpha_b = 12\cdot10^6$~K$^{-1}$ \cite{An-PRB2016}. 
Taking into account that spontaneous strains appear in BaTiO$_3$ below $T_C$, we obtain 
$u_{xx}(T)=a(T)a_0(T_g)^{-1}[1+\alpha_b(T-T_g)]^{-1}-1$ and $u_{yy}(T)=c(T)a_0(T_g)^{-1}[1+\alpha_b(T-T_g)]^{-1}-1$ for the a$_1$ domain, and \textit{vice versa} for the a$_2$ one.
Here $c$, $a$, and $a_0$ are the lattice constants of the tetragonal FE and cubic paraelectric phase of BaTiO$_3$, respectively.
We obtain $u_{xx}\approx-2.9\cdot10^{-3}$ and $u_{yy}\approx+7.9\cdot10^{-3}$ for the a$_1$ domain at RT by taking $c = 0.4035$\,nm, $a = 0.3992$\,nm \cite{BaTiO3-crystal}, and $a_0(T_g) = 0.4017$\,nm \cite{Kay-1949}. 

Additional in-plane anisotropy originating from the stripe shape of the domains has been shown to be negligible \cite{Lahtinen-AdvMat2013} and is not included in Eq.\,(\ref{eq:energy}).
The orientation of magnetization in the a$_1$ domain is defined by the total effective field, which is a sum of the external field $\mathbf{H}_\mathrm{ext}$, the effective magnetoelastic anisotropy field $\mathbf{H}_\mathrm{ME}$ and the out-of-plane effective field $\mathbf{H}_\mathrm{out}$:
\textcolor{newtext}{\begin{eqnarray}
    \mathbf{H}_\mathrm{eff}&=&-\frac{\partial F}{\mu_0\partial\mathbf{M}}=\mathbf{H}_\mathrm{ext}+\mathbf{H}_\mathrm{ME}+\mathbf{H}_\mathrm{out}\nonumber\\
    &=&\mathbf{H}_\mathrm{ext}-\frac{2B_1}{\mu_0M_S}\left(u_{xx}\mathbf{m}_{x}+u_{yy}\mathbf{m}_{y}\right)\nonumber\\
    &+&\left[-M_{S}+\frac{2B_1}{\mu_0M_S}(u_{xx}+u_{yy})\frac{c_{11}}{c_{12}}\right]\mathbf{m}_z,\label{Eq:Heff}
\end{eqnarray}}
where $\mathbf{m}_{i} = {m}_{i}\mathbf{e}_{i}$ with $\mathbf{e}_{i}$ being unit vectors directed along the coordinate axes.

\textcolor{newtext}{\subsection{Laser-induced change of magnetoelastic anisotropy}}

The light penetration depth in the CoFeB layer is below 20\,nm at the pump wavelength of 515\,nm, and, therefore, we argue that the anisotropy changes are caused by laser-induced processes in the FM film but not by changes in the FE substrate. 
Excitation of the metallic CoFeB film by a femtosecond laser pulse results in a rapid increase of the temperatures of electronic and ionic systems, which equilibrate after several picoseconds. 
This yields, first of all, ultrafast demagnetization, i.e., a subpicosecond decrease $\Delta M_S$ of the saturation magnetization \cite{Beaurepaire:PRL1996} followed by partial restoration upon equilibration between the ionic and electronic temperatures \cite{Koopmans:NatureMater2009}. 
Ultrafast demagnetization $\Delta M_S$ would increase the magnetoelastic anisotropy field $H_\mathrm{ME}\propto M_S^{-1}$. 
However, a rapid rise of the film temperature following excitation by a laser pulse should also lead to a decrease $\Delta B_1$ of the temperature-dependent magnetoelastic coupling parameter $B_1$.
In contrast to ultrafast demagnetization, this would reduce the effective magnetoelastic field $H_\mathrm{ME}\propto B_1$.
Finally, laser-induced heating $\Delta T$ modifies the substrate-induced film strains $u_{xx}$ and $u_{yy}$ via the term $\alpha_b\Delta T$ accounting for the thermal expansion of CoFeB, while the substrate temperature does not change significantly in our case.
The resulting variations $\Delta u_{xx}$ and $\Delta u_{yy}$ could alter the in-plane field $\mathbf{H}_\mathrm{ME}\propto (u_{xx}\mathbf{m}_x+u_{yy}\mathbf{m}_y)$ additionally.
As can be seen from Eq.\,\ref{Eq:Heff}, the changes $\Delta M_S$, $\Delta B_1$ and $\Delta  u_{xx(yy)}$ affect the out-of-plane anisotropy field $\mathbf{H}_\mathrm{out}$ and launch precession if $\mathbf{H}_\mathrm{out}$ is nonzero in the initial state \cite{vanKampen:PLR2002}, which is not the case in our experiments. 

\begin{figure}
\includegraphics[width=8.6cm]{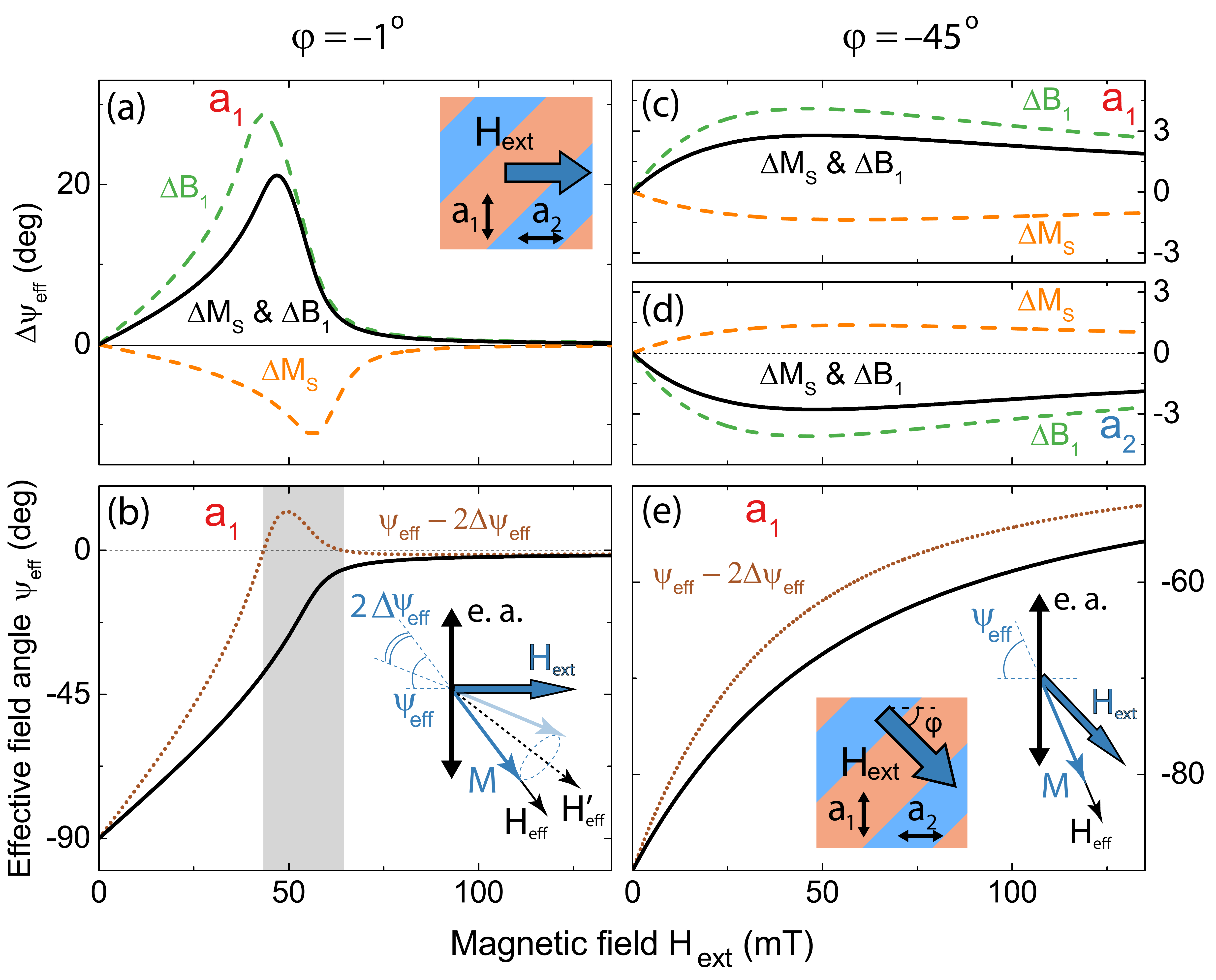}
\caption{(Color online) Calculated field dependencies of (a,b) the equilibrium effective field $\mathbf{H}_\mathrm{eff}$ orientation change $\Delta\psi_\mathrm{eff}$ in the a$_1$ (a,~c) and a$_2$ (d) domains due to laser-induced demagnetization $\Delta M_S$ (orange dashed line), the change of magnetoelastic parameter $\Delta B_1$ (green dashed line), and the total laser-induced effect resulting from $\Delta M_S$, $\Delta B_1$, and $\Delta u_{xx(yy)}$ (solid black line). (b,e) Calculated field dependence of the the equilibrium effective field $\mathbf{H}_\mathrm{eff}$ orientation $\Delta\psi_\mathrm{eff}$ (solid black line) and maximum magnetization deviation from $x$ axis in the a$_1$ domain $\psi_\mathrm{eff}-2\Delta\psi_\mathrm{eff}$ (dotted brown line). The external field $\mathbf{H_\mathrm{ext}}$ is applied (a,~b)
at $\varphi$~=~\ang{-1} and (c,~d,~e) at $\varphi$~=~\ang{-45}.} 
\label{Fig:Calc_inplane_Dev}
\end{figure}

In order to verify whether laser-induced changes of the in-plane magnetoelastic anisotropy can indeed account for the observed magnetization precession and to reveal which of the three contributions dominates, we performed numerical calculations.
The amplitude of the excited precession is defined by the \textcolor{newtext}{maximal} azimuthal angle $\Delta\psi_\mathrm{eff}$ by which the total effective field $\mathbf{H}_\mathrm{eff}$ reorients as a result of laser-induced changes $\Delta M_S$, $\Delta B_1$, and $\Delta u_{xx(yy)}$ [see inset in Fig.\,\ref{Fig:Calc_inplane_Dev}~(b,~e)].
In the experiments, out-of-plane oscillations of magnetization are detected, and their amplitude can be found as $\Delta M_z^0/M_S=\varsigma\Delta\psi_\mathrm{eff}$, where $\varsigma$ is the precession ellipticity  \cite{gurevich1996magnetization}.
\textcolor{newtext}{The precession frequency is governed by the external magnetic field, partially quenched magnetization, and modified magnetoelastic anisotropy within the laser excitation area.
At the timescale of the precession $\sim$1\,ns, the temperature within the excited spot, degree of the magnetization quenching and the magnetic anisotropy may vary.
However, if the changes in magnetic parameters of a medium are small, the Smit-Suhl formulae \cite{Smit:1955,Suhl-PRB:1955} can be used to calculate the precession frequency in a first-order approximation.}

\textcolor{newtext}{The degree of demagnetization was obtained experimentally.
Figure\,\ref{Fig:All_Measurements}~(b) shows the dynamics of ultrafast demagnetization $\Delta M_S(\Delta t)/M_S$ measured at the pump fluence $F=10$\,mJ$\cdot$cm$^{-2}$.
$\Delta M_S/M_S$ reaches 10\,\% within 2\,ps after excitation.
This is followed by slow restoration of magnetization.
After 1\,ns, the residual demagnetization is -5\,\%.}

Since a quasi-equilibrium between the electronic, ionic and spin systems establishes after several picoseconds following excitation, we assume that at $\Delta t>2$\,ps the laser-induced change of the magnetoelastic parameter $B_1$ relates to the demagnetization via the thermodynamic relation $B_1(T)/B_1(T=0)=\left[M_{S}(T)/M_S(T=0)\right]^{n(n+1)/2}$ \cite{Callen-JPCS1960}.
For the uniaxial magnetoelastic anisotropy of the individual domains, we use $n=2$ according to its single-ion origin \cite{Kittel:1960,Callen:PRB1965} and experimental data for Fe-Co-based amorphous alloys \cite{OHandley:SSComm1977,Barandiaran:PSS2011}.
\textcolor{newtext}{Then the laser-induced change of the magnetoelastic parameter are related to the demagnetization degree as
\begin{equation}
    \frac{\Delta B_1(\Delta t)}{B_1}\approx3\frac{\Delta M_S(\Delta t)}{M_S}+3\left(\frac{\Delta M_S(\Delta t)}{M_S}\right)^2.\label{Eq:ratio_dB1_dM}
\end{equation}
The ratio $\Delta B_1/B_1$ at a few picoseconds after excitation is estimated to be $-27$\% for $\Delta M_S/M_S=-10$\,\%. 
We note that the change of the magnetoelastic parameter is a nearly linear function of the degree of the demagnetization when the latter is within 0-15\,\%.}

\textcolor{newtext}{Estimation of the laser-induced strain change was done based on the laser-induced increase of the ionic temperature $\Delta T\approx$300\,K (see App.\,\ref{App:T} for details of temperature estimation). 
The strains in the a$_1$(a$_2$) domain change by $\Delta u_{xx}/u_{xx}\approx$124\,\% (-46\,\%) and $\Delta u_{yy}/u_{yy}\approx$-46\,\% (124\,\%).}

Figures\,\ref{Fig:Amps_Freqs} and \ref{Fig:Calc_inplane_Dev} summarize the results of the calculations for the a$_1$ and a$_2$ domains for the external magnetic field applied at $\varphi$~=~\ang{-1} and $\varphi$~=~\ang{-45} to the $x$\,axis.
A misalignment $\varphi$~=~\ang{-1} was introduced to achieve better agreement between the measured and calculated field dependencies of the precession frequency, and is within the precision of field alignment in the experiments.
An equilibrium RT magnetization $M_S=0.9\cdot10^{6}$\,A/m and magnetoelastic parameter $B_1=-22.8\cdot10^5$\,mJ/cm$^3$ were used to obtain good agreement between the calculated and experimental field dependencies of the precession frequency [Fig.\,\ref{Fig:Amps_Freqs}~(a,~b)].
The magnetoelastic anisotropy energy associated with the \ang{90} in-plane magnetization rotation $B_1(u_{yy}-u_{xx})=-2.445\cdot10^4$\,mJ/cm$^3$ and the magnetostriction coefficient $\lambda=14.5\cdot10^{-6}$ agree with previous works considering similar systems \cite{Brandl:2014,Isogami-pps2018}.

Figures\,\ref{Fig:Calc_inplane_Dev}(a,~c,~d) show the calculated field dependence of the change of the equilibrium angle $\psi_\mathrm{eff}$ between $\mathbf{H}_\mathrm{eff}$ and the $x$ axis under laser excitation.
Four scenarios are modeled, taking into account ultrafast demagnetization $\Delta M_s$ only (orange dashed line), the decrease of the magnetoelastic parameter $\Delta B_1$ only (green dashed line), the change of strains $\Delta u_{xx}$ and $\Delta u_{yy}$ only (not shown), and the combination of all three (solid black line).
As expected, $\Delta M_S$ and $\Delta B_1$ result in opposite signs of $\Delta\psi_\mathrm{eff}$.
The effect of $\Delta u_{xx}$ and $\Delta u_{yy}$ appears to be $>$20~times smaller than the one of $\Delta M_S$ and $\Delta B_1$, in agreement with previous studies \cite{Kats:PRB2016}.
Indeed, the thermal expansion reduces the tensile in-plane strain and increases the compressive one, thus affecting the magnetoelastic energy weakly. 
If all effects are combined, the laser-induced decrease of the magnetoelastic parameter $\Delta B_1$ dominates the magnetization response.
The reorientation of the total effective field is therefore dictated by a \textit{decrease} of the in-plane magnetoelastic anisotropy field $H_\mathrm{ME}$. 

In our model, the precession of magnetization is excited, i.e. $\Delta\psi_\mathrm{eff}\neq0$, in the whole studied field range in both domains at $\varphi~=$~\ang{-45} and only at low fields in the a$_1$ domain at $\varphi~=$~\ang{-1} [Figs.\,\ref{Fig:Calc_inplane_Dev}~(a,~c)].
The maximum absolute value of $\Delta\psi_\mathrm{eff}$ is close to \ang{20} if $\varphi=$~\ang{-1}, while it is several times smaller at $\varphi=$~\ang{-45}.
Taking into account the strong ellipticity $\varsigma$ of the excited precession, which depends on the applied field strength and orientation, we obtain
\textcolor{newtext}{ the field dependencies of the precession amplitudes $\Delta M_z^0/M_S$ shown in Fig.\,\ref{Fig:Amps_Freqs}~(c,~d) by solid lines.}

\textcolor{newtext}{In Fig.\,\ref{Fig:Amps_Freqs}~(c,~d) one sees that our model captures the general trend in the experimental field dependence of the precession amplitude.
At $\varphi=$~\ang{-45}, there is a reasonable agreement between experiment and model.
At  $\varphi=$~\ang{-1} both the experiment and the model show a pronounced maximum of the precession amplitude at low field and absence of the detectable precession at higher fields.
However, at $\varphi=$~\ang{-1} there are two discrepancies between the experimental and calculated amplitudes.
The first one is \textit{quantitative}, since the calculated amplitudes are larger then those obtained experimentally.
Partly, this is related to the uncertainty in the static polar Kerr rotation (see Sec.\,\ref{sec:experimental} and App.\,\ref{App:MOKE}).
This discrepancy can be compensated also by using a value of $n$ smaller than 2 in the relation $\Delta B_1/B_1=(\Delta M_S/M_S)^{n(n+1)/2}$, which is often the case in real uniaxial materials.}

\textcolor{newtext}{More important is the \textit{qualitative} difference between the experimental and calculated curves in Fig.\,\ref{Fig:Amps_Freqs}~(c). 
In our experiment the precession amplitude is maximum at 30\,mT, and no precession could be detected above a critical value of 50\,mT.
The model, in contrast, predicts that the maximum precession amplitude should be observed near the critical field $\sim50$\,mT.
Such behaviour of the excited precession amplitude has been indeed seen in previous experiments on laser-induced precession in a hard-axis configuration \cite{Kats:PRB2016}.
Note that varying the parameter ${n(n+1)/2}$ in the model does not change the characteristic field at which the maximum amplitude is observed.
Indeed, the considered excitation mechanism is nonresonant, and the maximum amplitude of the excited precession is achieved near a critical field where the magnetic system is most susceptible to any perturbation.
Achieving the maximum precession amplitude at a different field, and, correspondingly, at a different frequency, would require, for instance, resonant driving of the magnetic system at that frequency \cite{Jager:PRB2015}.
Neither the thicknesses of the Au and CoFeB layers, nor the periodicity of the domain pattern match this scenario. 
Below we argue that this \textit{qualitative} difference is, in fact, an indication of laser-induced magnetization switching.}

\textcolor{newtext}{\subsection{Magnetization dynamics triggered by the change of magnetoelastic anisotropy}}

\textcolor{newtext}{Excitation of the magnetization precession by laser pulses is one of the possible pathways for the magnetization switching \cite{Carpene:PRB2011,Stupakiewicz:Nature2017,Davies:PRL2019}.
In particular, it has been shown that ultrafast laser-driven decrease of magnetic anisotropy accompanied by the large-amplitude highly-damped precession enables magnetization switching in an applied magnetic field \cite{Davies:PRL2019}.
If the deviation $\Delta\psi_\mathrm{eff}$ of the effective field from its equilibrium direction towards applied magnetic field is large enough, and the in-plane magnetization component passes the applied field direction after one half of the period, the switching is initiated.
For such switching to be complete, the precession damping should be anomalously large to prevent the magnetization from returning to its initial orientation via further precessional motion \cite{Davies:PRL2019}.
Also, the restoration of the anisotropy should proceed at the timescale comparable with the precession frequency.
Once such switching occurs, the stroboscopic pump-probe technique fails to correctly measure the excited precession amplitude, since this technique requires the sample being in the very same initial state before excitation with each pump pulse.} 

\textcolor{newtext}{The abrupt drop of the pump-probe signal detected at $\varphi=$~\ang{-1} when the applied field reaches the critical value [Fig.\,\ref{Fig:Amps_Freqs}~(c)] may indicate that the excited precession amplitude is sufficiently large for precessional magnetization switching to be initiated.
In Fig.\,\ref{Fig:Calc_inplane_Dev} we illustrate this by plotting the maximum in-plane deviation of the magnetization from the $x$ axis when (b) $\varphi=$\ang{-1} and (e) $\varphi=$\ang{-45}.
Such a deviation is found as $\psi_\mathrm{eff}-2\Delta\psi_\mathrm{eff}$ (see inset in Fig.\,\ref{Fig:Calc_inplane_Dev}~(b,~e)).
Above $H_\mathrm{ext}\approx40$\,mT the first condition for the magnetization switching is satisfied for $\varphi=$\ang{-1} [see the grey area in Fig.\,\ref{Fig:Amps_Freqs}~(c)].
At $\varphi=$\ang{-45} this deviation is large at any external magnetic field and no switching is expected.}

\textcolor{newtext}{In order to analyze if such reorientation of the effective field can result in switching of magnetization, we simulated the trajectory of the magnetization motion in response to the abrupt decrease and following restoration of $M_S$ and $B_1$.
The simulations were performed by solving the Landau-Lifshitz-Gilbert equation for a macrospin.
External field is applied along the hard magnetization axis ($x-$axis).
The time-dependent effective field entering the Landau-Lifshitz-Gilbert equation is calculated using Eq.\,(\ref{Eq:Heff}) and taking into account the relaxation of the laser-induced demagnetization found in our experiments [see Fig.\,\ref{Fig:All_Measurements}~(b) and Fig.\,\ref{Fig:Model_switching}~(a)].
The temporal evolution of $\Delta B_1$ is found accordingly using Eq.\,(\ref{Eq:ratio_dB1_dM}).
Also, slower relaxation of $M_S$ and $B_1$ towards equilibrium is considered, as shown in Fig.\,\ref{Fig:Model_switching}~(b).
The Gilbert damping parameter $\alpha=0.02$ is chosen to match the precession decay observed far from the critical field.}

\textcolor{newtext}{\begin{figure}
\includegraphics[width=8.6cm]{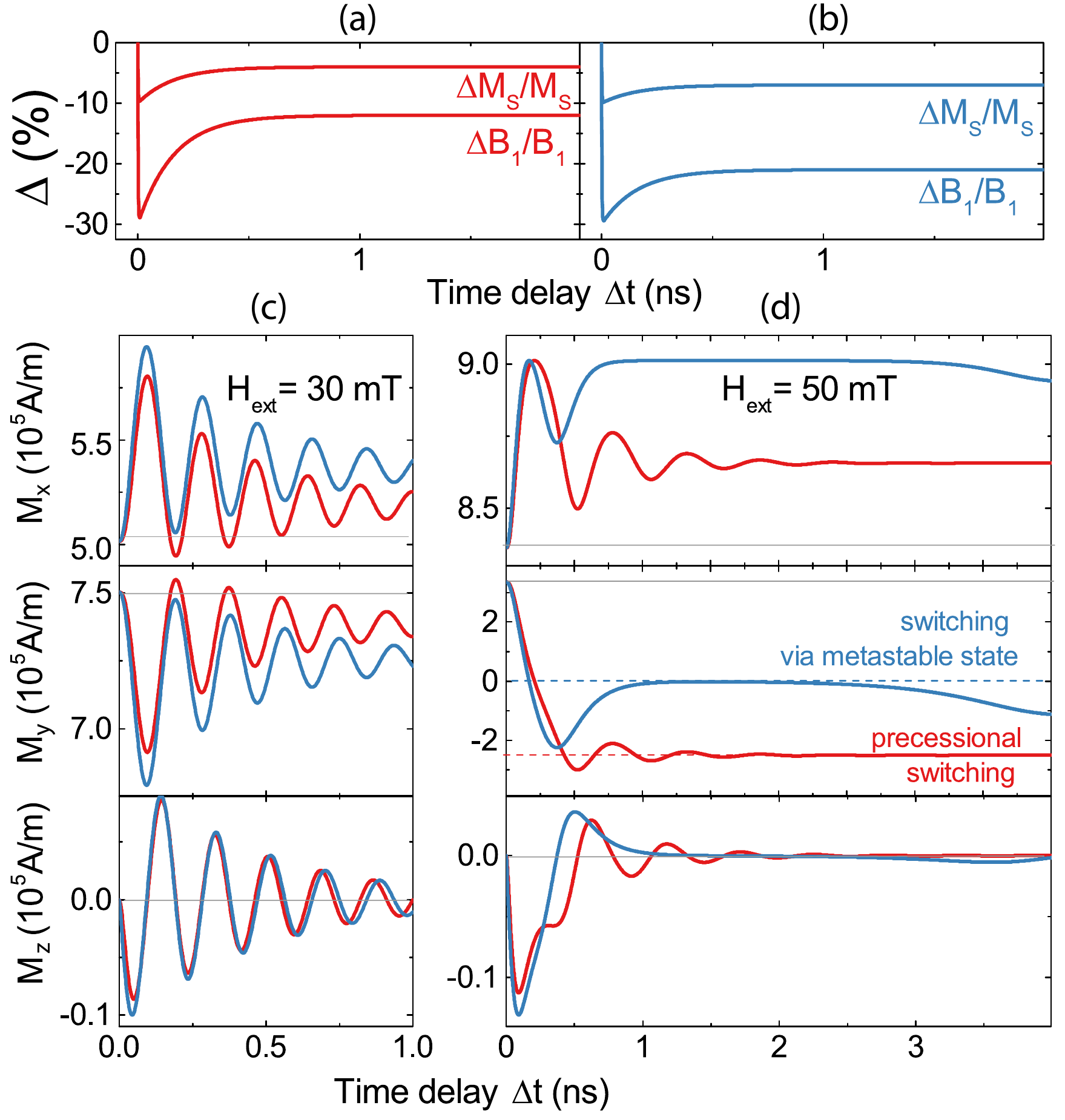}
\caption{\textcolor{newtext}{(Color online) Modeling of the laser-induced magnetization precession when the magnetic field $\mathbf{H}_\mathrm{ext}$ is applied along the hard ($x$) axis. (a,b) Evolution of the magnetization $M_S$ and magnetoelastic parameter $B_1$ used in the modeling, with relaxation (a) similar to that observed in the experiment [Fig.\,\ref{Fig:All_Measurements}~(b)] and (b) slower one. (c,d) Dynamics of $M_{x,y,z}$ components of magnetization at (c) $H_\mathrm{ext}=30$\,mT and (d) $H_\mathrm{ext}=50$\,mT. Results are obtained by using the faster (red lines) and slower (blue lines) evolution of the $M_S$ and $B_1$ shown in the panels (a) and (b), respectively. Solid grey lines indicate the equilibrium orientation of the magnetization. Red and blue dashed lines indicate the switched state and metastable state ($\mathbf{M}\|\mathbf{H}_\mathrm{ext}$), respectively.} } 
\label{Fig:Model_switching}
\end{figure}}

\textcolor{newtext}{In Fig.\,\ref{Fig:Model_switching} we show the simulated magnetization trajectories represented by evolution of the $x-,y-,z-$ components of magnetization at $H_\mathrm{ext}=$~30 and 50\,mT.
As one can see in  Fig.\,\ref{Fig:Model_switching}~(c), at $H_\mathrm{ext}=$~30~mT laser excitation results in decaying magnetization precession.
Changing the details of the relaxation of $M_S$ and $B_1$ towards equilibrium has only little effect on the excited dynamics [compare red and blue curves in Fig.\,\ref{Fig:Model_switching}(c)].
At $H_\mathrm{ext}=$~50~mT [Fig.\,\ref{Fig:Model_switching}~(d)] the amplitude of the excited precession is large enough and the condition for the initiation of precessional switching is fulfilled.
The subsequent dynamics is strongly dependent on the relaxation of the total effective field towards its equilibrium, and we get either precessional switching [red lines in Fig.\,\ref{Fig:Model_switching}(d)] or the temporal trapping of magnetization at metastable local states.
The subsequent cooling down of the system results in either a return of the magnetization to its initial orientation, or magnetic switching to another stable direction [blue lines in Fig.\,\ref{Fig:Model_switching}(d)].
This leads to a situation where only part of the excitation events result in switching.}

\textcolor{newtext}{Importantly, once the magnetization has switched from one equilibrium orientation to another, the next pump pulse would again excite the precession and switching.
However, the $M_z$ component would oscillate with a $\pi-$shift in phase as compared to the oscillations excited by the preceeding pump pulse. 
Our pump-probe experiments are stroboscopic and average the dynamics of $M_z$ which is repetitively driven by $\sim$10$^4$ pump pulses.
Therefore, the described switching is seen as a reduction of the detected signal amplitude.
At 50\% switching events no oscillatory signal would be detected.
Furthermore, the Gaussian spatial profile of the pump pulse causes switching in the center of the spot at lower average fluence, as compared to the outer part.
This, in particular, should result in concentric ring-like structures with switched and non-switched areas, which is a characteristic feature of the precessional switching driven by spatially inhomogeneous stimuli \cite{Tudosa:Nature2004,deJong:PRL2012,Davies:PRL2019}. 
Detection of such switching requires single-shot pump-probe imaging \cite{KIMEL2020}, for which the strength of the magneto-optical signal in CoFeB/BaTiO$_3$ is insufficient.
Nevertheless, we argue that the noticeable decrease of the detected amplitude of the precession at the applied fields exceeding 30\,mT seen in our experiments is a manifestation of precessional switching of magnetization triggered by an ultrafast decrease of the magnetoelastic parameter.}

\section{Conclusions}

In conclusion, we have shown that the magnetoelastic coupling in a metallic CoFeB film can be significantly reduced on a picosecond timescale by excitation with a femtosecond laser pulse.
This effect is explained by a simple model, which accounts for a laser-induced increase of electronic and ionic temperatures and relates the induced changes of the magnetoelastic parameter and magnetization via a $n(n+1)/2$-law. 
This law appears to hold for laser-induced processes in metals at timescales beyond several picoseconds following excitation, which is required for establishing a quasi-equilibrium between the electronic, ionic and spin subsystems.
The demonstrated ability to decrease the magnetoelastic parameter by laser pulses enables the driving of magnetization dynamics in metallic films wherein the magnetoelastic anisotropy dominates.
We employed this mechanism to experimentally realize selective excitation of magnetization precession in individual micron-size magnetic stripe domains imprinted in a CoFeB film by a ferroelectric BaTiO$_3$ substrate.
We found that the excitation of precession in an individual CoFeB domain is controlled by the strength and the orientation of the applied magnetic field with respect to the local uniaxial anisotropy axis. This allows, in particular, distinguishing magnetization precessions in separate domains even when they are unresolvable in static measurements due to large applied magnetic field. 
\textcolor{newtext}{Further, we show that the change of the magnetoelastic anisotropy even triggers the precessional switching of the magnetization in a selected domain, which manifests itself in the pump-probe experiments as a suppression of the detected precession. Precessional switching is realized near a critical ﬁeld of 50\,mT applied along the hard magnetization axis of the domain.}

\textcolor{newtext}{Our results therefore show that structures consisting of CoFeB on ferroelectric or piezoelectric substrates are prospective candidates for optically-driven precessional switching via anomalously-damped precession \cite{Davies:PRL2019} under application-relevant conditions. 
Indeed, speciﬁc anisotropy and damping required for the switching may be tailored in such heterostructures at both the growth stage and afterwards. Namely, ferroelectric or piezoelectric substrate enables controlling the orientation and strength of magnetic anisotropy by external voltage, while the damping in CoFeB may be tuned by various means, such as annealing and composition \cite{Zhang:APL2011,Srivastava:PRApp2018}.}

\textcolor{newtext}{The possibility to alter the magnetoelastic parameter by laser pulses can be exploited further in both FM-FE composites and in heterostructures formed by FM and piezoelectric constituents. A decrease of $B_1$ reduces the strain-mediated magnetoelectric coupling in multiferroic and magnetoelectric heterostructures, which could enable laser-assisted magnetization switching by applied electric field with submicrometer spatial resolution \cite{LoContePRMater:2018}, in an analogy to the heat-assisted magnetic recording. Moreover, in strained ﬁlms a spin reorientation transition can occur due to an extra contribution to the out-of-plane anisotropy [Eq.\,\ref{eq:energy}] controlled by the magnetoelastic parameter \cite{Pertsev-PRB2008}. Hence, laser excitation may be used to trigger such a transition at picosecond timescale via changes of $B_1$ and ﬁlm strains.
Finally, laser-induced changes of magnetoelastic anisotropy may be employed for driving spin waves \cite{Khokhlov:2018} in switchable magnonic waveguides based on CoFeB/BaTiO$_3$ \cite{Brandl:2014,VanDeWiele:2016,LopezGonzalez:2016,HamalainenNature:2018}.}

\section{Acknowlegments}
We thank N.\,E.\,Khokhlov and P.\,I.\,Gerevenkov for the fruitful discussions.
TEM study was carried out using Jeol JEM-2100F microscope of the Federal Joint Research Center "Materials science and characterization in advanced technologies".
L.A.Sh., A.V.Sch. and A.M.K. acknowledge RSF (grant No.\,16-12-10485) for the support of the experimental studies, and the RFBR (grant No.\,19-52-12065) and the DFG programme TRR160 ICRC for the support of the theoretical work. LLG-based modeling was performed by L.A.Sh. and A.M.K. under support of the RSF (Grant No. 20-12-00309). 
S.J.H. and S.v.D. were supported by the Academy of Finland (grant No.\,316857). The collaboration between the Ioffe Institute and Aalto University is a part of the COST Action CA17123 Magnetofon.

\appendix
\section{Static polar and longitudinal MOKE\\ in CoFeB/BaTiO$_3$}\label{App:MOKE}
\textcolor{newtext}{In experiment we have measured static longitudinal Kerr rotation $\theta^L_S$ and ellipticity $\epsilon^L_S$ for the in-plane saturated sample. Using the general formulae for the longitudinal Kerr effect \cite{Zvezdin_book} we find the magneto-optical parameter $Q$ as:
\begin{equation}\label{Eq:Q_MO_parameter}
Q = (i\cdot\theta^L_S + \epsilon^L_S)/A_L,
\end{equation}
where 
\begin{equation*}
A_L = \frac{\eta^2\sin\zeta(\sin\zeta\tan\zeta - \sqrt{\eta^2 - \sin^2\zeta})}{(\eta^2 - 1)(\eta^2 - \tan^2\zeta)\sqrt{\eta^2 - \sin^2\zeta}}.
\end{equation*}
$\zeta=$~\ang{45} is the angle of incidence used in the longitudinal MOKE experiment. $\eta = n_2/n_1$, $n_1$
and $n_2$ are refractive indices of gold and CoFeB at the probe wavelength of 1030\,nm, respectively.}

\textcolor{newtext}{Knowing $Q$ one can calculate the polar Kerr rotation $\theta^P_S$ corresponding to the fully saturated out of plane film as:
\begin{equation}\label{Eq:Polar_rotation}
\theta^P_S = \mathrm{Im}\{Q A_P\},
\end{equation}
where 
\begin{equation*}
A_P = \frac{\eta^2(\sin\zeta\tan\zeta + \sqrt{\eta^2 - \sin^2\zeta})}{(\eta^2 - 1)(\eta^2 - \tan^2\zeta)}.
\end{equation*}
Here $\zeta=$~\ang{17} is the angle of incidence used in the main experiments.
Coefficients $A_L$ and $A_P$ are written for $p$-polarization.}

\textcolor{newtext}{We are not aware of the data on refractive index for Co$_{40}$Fe$_{40}$B$_{20}$ at 1030\,nm. In \cite{Liang:OSA2015} the refractive index $n_2=4+i4.7$ for the Co$_{20}$Fe$_{60}$B$_{20}$ films at 1030\,nm has been reported. Comparison of the spectra of complex refractive indices of as-deposited  Co$_{20}$Fe$_{60}$B$_{20}$ \cite{Liang:OSA2015} and Co$_{60}$Fe$_{20}$B$_{20}$ \cite{Hoffmann:2019} for the 400-800\,nm wavelengths suggests that the optical properties of the alloy do not vary strongly within this composition range. However, optical properties of CoFeB films depend on the crystallinity of the film \cite{Hoffmann:2019}. Furthermore, the optical parameters of thin gold layers are dependent on substrates, thickness etc. \cite{Rosenblatt:PRX2020}.} 

\textcolor{newtext}{Using $n_2=4+i4.7$ for CoFeB and $n_1 = 1.5 + i5.3$ reported for an 11\,nm thick gold film \cite{Rosenblatt:PRX2020}, we get an estimate for the polar Kerr rotation $\theta^P_S\approx15$\,mdeg. Taking different optical parameters of CoFeB and Au may increase or decrease estimated $\theta^P_S$.}

\section{Laser-induced heating of CoFeB layer}\label{App:T}
\textcolor{newtext}{For the estimation of ultrafast laser-induced heating of the CoFeB layer in Au/CoFeB/\ch{BaTiO3} we used the following parameters.
The complex refractive index of the capping gold layer is $n_1 = 0.51 + i2.02$ at the pump wavelength 515\,nm \cite{Rosenblatt:PRX2020}. This gives intensity reflection coefficient $R_1=$~0.68 at the interface air/Au. Absorption coefficient of the gold film is $\alpha_1=$~4.28$\cdot10^5$~cm$^{-1}$.
The complex refractive index of the CoFeB is $n_2 = 2.5 + i3.25$ at 515\,nm \cite{Liang:OSA2015,Hoffmann:2019} which gives the intensity reflection coefficient at the Au/CoFeB interface  $R_2=$~0.148.
Absorption coefficient of CoFeB $\alpha_2=$~7.93$\cdot10^5$~cm$^{-1}$.}

\textcolor{newtext}{The temperature increase for the CoFeB layer is found as:
\begin{equation}
\Delta T=(1-R_1)e^{-\alpha_1d}(1-R_2)\alpha_2(1-e^{-1})\frac{J}{C \rho},
\end{equation}
where $C=$~440\,J~kg$^{-1}$K$^{-1}$ is heat capacity \cite{walter:natMat2011} and $\rho=$~7.7$\cdot10^3$~kg~m$^{-3}$ is density of CoFeB \cite{oHandley:APL1976}, $d$ is thickness of Au layer. The factor $\alpha_2(1-e^{-1})J$ gives the volume energy density in the case of the film thickness exceeding the light penetration depth \cite{walowski2012PhD}.
For the incident fluence of $J=10$\,mJ/cm$^{2}$ one gets laser-induced heating of $\Delta T\sim$300\,K.
Accounting for the additional scattering of light which takes place at the Au/CoFeB interface may reduce the amount of energy deposited in CoFeB and the corresponding heating.}

\bibliography{References}

\begin{thebibliography}{76}%
\makeatletter
\providecommand \@ifxundefined [1]{%
 \@ifx{#1\undefined}
}%
\providecommand \@ifnum [1]{%
 \ifnum #1\expandafter \@firstoftwo
 \else \expandafter \@secondoftwo
 \fi
}%
\providecommand \@ifx [1]{%
 \ifx #1\expandafter \@firstoftwo
 \else \expandafter \@secondoftwo
 \fi
}%
\providecommand \natexlab [1]{#1}%
\providecommand \enquote  [1]{``#1''}%
\providecommand \bibnamefont  [1]{#1}%
\providecommand \bibfnamefont [1]{#1}%
\providecommand \citenamefont [1]{#1}%
\providecommand \href@noop [0]{\@secondoftwo}%
\providecommand \href [0]{\begingroup \@sanitize@url \@href}%
\providecommand \@href[1]{\@@startlink{#1}\@@href}%
\providecommand \@@href[1]{\endgroup#1\@@endlink}%
\providecommand \@sanitize@url [0]{\catcode `\\12\catcode `\$12\catcode
  `\&12\catcode `\#12\catcode `\^12\catcode `\_12\catcode `\%12\relax}%
\providecommand \@@startlink[1]{}%
\providecommand \@@endlink[0]{}%
\providecommand \url  [0]{\begingroup\@sanitize@url \@url }%
\providecommand \@url [1]{\endgroup\@href {#1}{\urlprefix }}%
\providecommand \urlprefix  [0]{URL }%
\providecommand \Eprint [0]{\href }%
\providecommand \doibase [0]{http://dx.doi.org/}%
\providecommand \selectlanguage [0]{\@gobble}%
\providecommand \bibinfo  [0]{\@secondoftwo}%
\providecommand \bibfield  [0]{\@secondoftwo}%
\providecommand \translation [1]{[#1]}%
\providecommand \BibitemOpen [0]{}%
\providecommand \bibitemStop [0]{}%
\providecommand \bibitemNoStop [0]{.\EOS\space}%
\providecommand \EOS [0]{\spacefactor3000\relax}%
\providecommand \BibitemShut  [1]{\csname bibitem#1\endcsname}%
\let\auto@bib@innerbib\@empty
\bibitem [{\citenamefont {Smolenski\v{i}}\ and\ \citenamefont
  {Chupis}(1982)}]{Smolenskii:1982}%
  \BibitemOpen
  \bibfield  {author} {\bibinfo {author} {\bibfnamefont {G.~A.}\ \bibnamefont
  {Smolenski\v{i}}}\ and\ \bibinfo {author} {\bibfnamefont {I.~E.}\
  \bibnamefont {Chupis}},\ }\bibfield  {title} {\enquote {\bibinfo {title}
  {Ferroelectromagnets},}\ }\href {\doibase
  https://doi.org/10.1070/PU1982v025n07ABEH004570} {\bibfield  {journal}
  {\bibinfo  {journal} {Sov. Phys.: Uspekhi}\ }\textbf {\bibinfo {volume}
  {25}},\ \bibinfo {pages} {475--493} (\bibinfo {year} {1982})}\BibitemShut
  {NoStop}%
\bibitem [{\citenamefont {Fiebig}(2005)}]{Fiebig:2005}%
  \BibitemOpen
  \bibfield  {author} {\bibinfo {author} {\bibfnamefont {M.}~\bibnamefont
  {Fiebig}},\ }\bibfield  {title} {\enquote {\bibinfo {title} {Revival of the
  magnetoelectric effect},}\ }\href {\doibase 10.1088/0022-3727/38/8/r01}
  {\bibfield  {journal} {\bibinfo  {journal} {J. Phys. D: Appl. Phys.}\
  }\textbf {\bibinfo {volume} {38}},\ \bibinfo {pages} {R123--R152} (\bibinfo
  {year} {2005})}\BibitemShut {NoStop}%
\bibitem [{\citenamefont {Hu}\ \emph {et~al.}(2015)\citenamefont {Hu},
  \citenamefont {Chen},\ and\ \citenamefont {Nan}}]{Hu:2015}%
  \BibitemOpen
  \bibfield  {author} {\bibinfo {author} {\bibfnamefont {J.-M.}\ \bibnamefont
  {Hu}}, \bibinfo {author} {\bibfnamefont {L.-Q.}\ \bibnamefont {Chen}}, \ and\
  \bibinfo {author} {\bibfnamefont {C.-W.}\ \bibnamefont {Nan}},\ }\bibfield
  {title} {\enquote {\bibinfo {title} {Multiferroic heterostructures
  integrating ferroelectric and magnetic materials},}\ }\href {\doibase
  10.1002/adma.201502824} {\bibfield  {journal} {\bibinfo  {journal} {Adv.
  Mater.}\ }\textbf {\bibinfo {volume} {28}},\ \bibinfo {pages} {15--39}
  (\bibinfo {year} {2015})}\BibitemShut {NoStop}%
\bibitem [{\citenamefont {Manipatruni}\ \emph {et~al.}(2019)\citenamefont
  {Manipatruni}, \citenamefont {Nikonov}, \citenamefont {Lin}, \citenamefont
  {Gosavi}, \citenamefont {Liu}, \citenamefont {Prasad}, \citenamefont {Huang},
  \citenamefont {Bonturim}, \citenamefont {Ramesh},\ and\ \citenamefont
  {Young}}]{Manipatruni-Nature2019}%
  \BibitemOpen
  \bibfield  {author} {\bibinfo {author} {\bibfnamefont {S.}~\bibnamefont
  {Manipatruni}}, \bibinfo {author} {\bibfnamefont {D.~E.}\ \bibnamefont
  {Nikonov}}, \bibinfo {author} {\bibfnamefont {C.-C.}\ \bibnamefont {Lin}},
  \bibinfo {author} {\bibfnamefont {T.~A.}\ \bibnamefont {Gosavi}}, \bibinfo
  {author} {\bibfnamefont {H.}~\bibnamefont {Liu}}, \bibinfo {author}
  {\bibfnamefont {B.}~\bibnamefont {Prasad}}, \bibinfo {author} {\bibfnamefont
  {Y.-L.}\ \bibnamefont {Huang}}, \bibinfo {author} {\bibfnamefont
  {E.}~\bibnamefont {Bonturim}}, \bibinfo {author} {\bibfnamefont
  {R.}~\bibnamefont {Ramesh}}, \ and\ \bibinfo {author} {\bibfnamefont {I.~A.}\
  \bibnamefont {Young}},\ }\bibfield  {title} {\enquote {\bibinfo {title}
  {Scalable energy-efficient magnetoelectric spin–orbit logic},}\ }\href
  {\doibase 10.1038/s41586-018-0770-2} {\bibfield  {journal} {\bibinfo
  {journal} {Nature}\ }\textbf {\bibinfo {volume} {565}},\ \bibinfo {pages}
  {35} (\bibinfo {year} {2019})}\BibitemShut {NoStop}%
\bibitem [{\citenamefont {Chu}\ \emph {et~al.}(2018)\citenamefont {Chu},
  \citenamefont {PourhosseiniAsl},\ and\ \citenamefont {Dong}}]{Chu-JPD2018}%
  \BibitemOpen
  \bibfield  {author} {\bibinfo {author} {\bibfnamefont {Z.}~\bibnamefont
  {Chu}}, \bibinfo {author} {\bibfnamefont {M.}~\bibnamefont
  {PourhosseiniAsl}}, \ and\ \bibinfo {author} {\bibfnamefont {S.}~\bibnamefont
  {Dong}},\ }\bibfield  {title} {\enquote {\bibinfo {title} {Review of
  multi-layered magnetoelectric composite materials and devices
  applications},}\ }\href {\doibase https://doi.org/10.1088/1361-6463/aac29b}
  {\bibfield  {journal} {\bibinfo  {journal} {J. Phys. D: Appl. Phys.}\
  }\textbf {\bibinfo {volume} {51}},\ \bibinfo {pages} {243001} (\bibinfo
  {year} {2018})}\BibitemShut {NoStop}%
\bibitem [{\citenamefont {Hill}(2000)}]{Hill-JPCB2000}%
  \BibitemOpen
  \bibfield  {author} {\bibinfo {author} {\bibfnamefont {N.~A.}\ \bibnamefont
  {Hill}},\ }\bibfield  {title} {\enquote {\bibinfo {title} {Why are there so
  few magnetic ferroelectrics?}}\ }\href {\doibase 10.1021/jp000114x}
  {\bibfield  {journal} {\bibinfo  {journal} {J. Phys. Chem. B}\ }\textbf
  {\bibinfo {volume} {104}},\ \bibinfo {pages} {6694--6709} (\bibinfo {year}
  {2000})}\BibitemShut {NoStop}%
\bibitem [{\citenamefont {Vaz}(2012)}]{Vaz:2012}%
  \BibitemOpen
  \bibfield  {author} {\bibinfo {author} {\bibfnamefont {C.~A.~F.}\
  \bibnamefont {Vaz}},\ }\bibfield  {title} {\enquote {\bibinfo {title}
  {Electric field control of magnetism in multiferroic heterostructures},}\
  }\href {\doibase https://doi.org/10.1088/0953-8984/24/33/333201} {\bibfield
  {journal} {\bibinfo  {journal} {J. Phys. Condens. Matter}\ }\textbf {\bibinfo
  {volume} {24}},\ \bibinfo {pages} {333201} (\bibinfo {year}
  {2012})}\BibitemShut {NoStop}%
\bibitem [{\citenamefont {Wang}\ \emph {et~al.}(2010)\citenamefont {Wang},
  \citenamefont {Hu}, \citenamefont {Lin},\ and\ \citenamefont
  {Nan}}]{Wang-NPGAsia2010}%
  \BibitemOpen
  \bibfield  {author} {\bibinfo {author} {\bibfnamefont {Y.}~\bibnamefont
  {Wang}}, \bibinfo {author} {\bibfnamefont {J.}~\bibnamefont {Hu}}, \bibinfo
  {author} {\bibfnamefont {Y.}~\bibnamefont {Lin}}, \ and\ \bibinfo {author}
  {\bibfnamefont {C.-W.}\ \bibnamefont {Nan}},\ }\bibfield  {title} {\enquote
  {\bibinfo {title} {Multiferroic magnetoelectric composite nanostructures},}\
  }\href {\doibase 10.1038/asiamat.2010.32} {\bibfield  {journal} {\bibinfo
  {journal} {NPG Asia Mater.}\ }\textbf {\bibinfo {volume} {2}},\ \bibinfo
  {pages} {61--68} (\bibinfo {year} {2010})}\BibitemShut {NoStop}%
\bibitem [{\citenamefont {Carman}\ and\ \citenamefont
  {Sun}(2018)}]{carman-MRS2018}%
  \BibitemOpen
  \bibfield  {author} {\bibinfo {author} {\bibfnamefont {G.~P.}\ \bibnamefont
  {Carman}}\ and\ \bibinfo {author} {\bibfnamefont {N.}~\bibnamefont {Sun}},\
  }\bibfield  {title} {\enquote {\bibinfo {title} {Strain-mediated
  magnetoelectrics: Turning science fiction into reality},}\ }\href {\doibase
  https://doi.org/10.1557/mrs.2018.236} {\bibfield  {journal} {\bibinfo
  {journal} {MRS Bull.}\ }\textbf {\bibinfo {volume} {43}},\ \bibinfo {pages}
  {822--828} (\bibinfo {year} {2018})}\BibitemShut {NoStop}%
\bibitem [{\citenamefont {Nan}\ \emph {et~al.}(2008)\citenamefont {Nan},
  \citenamefont {Bichurin}, \citenamefont {Dong}, \citenamefont {Viehland},\
  and\ \citenamefont {Srinivasan}}]{Nan:JAP2008}%
  \BibitemOpen
  \bibfield  {author} {\bibinfo {author} {\bibfnamefont {C.-W.}\ \bibnamefont
  {Nan}}, \bibinfo {author} {\bibfnamefont {M.~I.}\ \bibnamefont {Bichurin}},
  \bibinfo {author} {\bibfnamefont {S.}~\bibnamefont {Dong}}, \bibinfo {author}
  {\bibfnamefont {D.}~\bibnamefont {Viehland}}, \ and\ \bibinfo {author}
  {\bibfnamefont {G.}~\bibnamefont {Srinivasan}},\ }\bibfield  {title}
  {\enquote {\bibinfo {title} {Multiferroic magnetoelectric composites:
  Historical perspective, status, and future directions},}\ }\href {\doibase
  10.1063/1.2836410} {\bibfield  {journal} {\bibinfo  {journal} {J. Appl.
  Phys.}\ }\textbf {\bibinfo {volume} {103}},\ \bibinfo {pages} {031101}
  (\bibinfo {year} {2008})}\BibitemShut {NoStop}%
\bibitem [{\citenamefont {Taniyama}\ \emph {et~al.}(2007)\citenamefont
  {Taniyama}, \citenamefont {Akasaka}, \citenamefont {Fu}, \citenamefont
  {Itoh}, \citenamefont {Takashima},\ and\ \citenamefont
  {Prijamboedi}}]{Taniyama-JAP2007}%
  \BibitemOpen
  \bibfield  {author} {\bibinfo {author} {\bibfnamefont {T.}~\bibnamefont
  {Taniyama}}, \bibinfo {author} {\bibfnamefont {K.}~\bibnamefont {Akasaka}},
  \bibinfo {author} {\bibfnamefont {D.}~\bibnamefont {Fu}}, \bibinfo {author}
  {\bibfnamefont {M.}~\bibnamefont {Itoh}}, \bibinfo {author} {\bibfnamefont
  {H.}~\bibnamefont {Takashima}}, \ and\ \bibinfo {author} {\bibfnamefont
  {B.}~\bibnamefont {Prijamboedi}},\ }\bibfield  {title} {\enquote {\bibinfo
  {title} {Electrical voltage manipulation of ferromagnetic microdomain
  structures in a ferromagnetic/ferroelectric hybrid structure},}\ }\href
  {\doibase 10.1063/1.2711280} {\bibfield  {journal} {\bibinfo  {journal} {J.
  Appl. Phys.}\ }\textbf {\bibinfo {volume} {101}},\ \bibinfo {pages} {09F512}
  (\bibinfo {year} {2007})}\BibitemShut {NoStop}%
\bibitem [{\citenamefont {Lahtinen}\ \emph {et~al.}(2013)\citenamefont
  {Lahtinen}, \citenamefont {Tuomi},\ and\ \citenamefont {van
  Dijken}}]{Lahtinen-AdvMat2013}%
  \BibitemOpen
  \bibfield  {author} {\bibinfo {author} {\bibfnamefont {T.~H.~E.}\
  \bibnamefont {Lahtinen}}, \bibinfo {author} {\bibfnamefont {J.~O.}\
  \bibnamefont {Tuomi}}, \ and\ \bibinfo {author} {\bibfnamefont
  {S.}~\bibnamefont {van Dijken}},\ }\bibfield  {title} {\enquote {\bibinfo
  {title} {Pattern transfer and electric-field-induced magnetic domain
  formation in multiferroic heterostructures},}\ }\href {\doibase
  10.1002/adma.201100426} {\bibfield  {journal} {\bibinfo  {journal} {Adv.
  Mater.}\ }\textbf {\bibinfo {volume} {23}},\ \bibinfo {pages} {3187--3191}
  (\bibinfo {year} {2013})}\BibitemShut {NoStop}%
\bibitem [{\citenamefont {Shirahata}\ \emph {et~al.}(2015)\citenamefont
  {Shirahata}, \citenamefont {Shiina}, \citenamefont {Gonz{\'a}lez},
  \citenamefont {Franke}, \citenamefont {Wada}, \citenamefont {Itoh},
  \citenamefont {Pertsev}, \citenamefont {van Dijken},\ and\ \citenamefont
  {Taniyama}}]{shirahata-NPGAsia2015}%
  \BibitemOpen
  \bibfield  {author} {\bibinfo {author} {\bibfnamefont {Y.}~\bibnamefont
  {Shirahata}}, \bibinfo {author} {\bibfnamefont {R.}~\bibnamefont {Shiina}},
  \bibinfo {author} {\bibfnamefont {D.~L.}\ \bibnamefont {Gonz{\'a}lez}},
  \bibinfo {author} {\bibfnamefont {K.~J.~A.}\ \bibnamefont {Franke}}, \bibinfo
  {author} {\bibfnamefont {E.}~\bibnamefont {Wada}}, \bibinfo {author}
  {\bibfnamefont {M.}~\bibnamefont {Itoh}}, \bibinfo {author} {\bibfnamefont
  {N.~A.}\ \bibnamefont {Pertsev}}, \bibinfo {author} {\bibfnamefont
  {S.}~\bibnamefont {van Dijken}}, \ and\ \bibinfo {author} {\bibfnamefont
  {T.}~\bibnamefont {Taniyama}},\ }\bibfield  {title} {\enquote {\bibinfo
  {title} {Electric-field switching of perpendicularly magnetized
  multilayers},}\ }\href {\doibase https://doi.org/10.1038/am.2015.72}
  {\bibfield  {journal} {\bibinfo  {journal} {NPG Asia Mater.}\ }\textbf
  {\bibinfo {volume} {7}},\ \bibinfo {pages} {e198} (\bibinfo {year}
  {2015})}\BibitemShut {NoStop}%
\bibitem [{\citenamefont {Lahtinen}\ \emph {et~al.}(2012)\citenamefont
  {Lahtinen}, \citenamefont {Franke},\ and\ \citenamefont
  {Van~Dijken}}]{lahtinen-SciRep2012}%
  \BibitemOpen
  \bibfield  {author} {\bibinfo {author} {\bibfnamefont {T.~H.~E.}\
  \bibnamefont {Lahtinen}}, \bibinfo {author} {\bibfnamefont {K.~J.~A.}\
  \bibnamefont {Franke}}, \ and\ \bibinfo {author} {\bibfnamefont
  {S.}~\bibnamefont {Van~Dijken}},\ }\bibfield  {title} {\enquote {\bibinfo
  {title} {Electric-field control of magnetic domain wall motion and local
  magnetization reversal},}\ }\href {\doibase
  https://doi.org/10.1038/srep00258} {\bibfield  {journal} {\bibinfo  {journal}
  {Sci. Rep.}\ }\textbf {\bibinfo {volume} {2}},\ \bibinfo {pages} {258}
  (\bibinfo {year} {2012})}\BibitemShut {NoStop}%
\bibitem [{\citenamefont {Buzzi}\ \emph {et~al.}(2013)\citenamefont {Buzzi},
  \citenamefont {Chopdekar}, \citenamefont {Hockel}, \citenamefont {Bur},
  \citenamefont {Wu}, \citenamefont {Pilet}, \citenamefont {Warnicke},
  \citenamefont {Carman}, \citenamefont {Heyderman},\ and\ \citenamefont
  {Nolting}}]{buzzi-PRL2013}%
  \BibitemOpen
  \bibfield  {author} {\bibinfo {author} {\bibfnamefont {M.}~\bibnamefont
  {Buzzi}}, \bibinfo {author} {\bibfnamefont {R.~V.}\ \bibnamefont
  {Chopdekar}}, \bibinfo {author} {\bibfnamefont {J.~L.}\ \bibnamefont
  {Hockel}}, \bibinfo {author} {\bibfnamefont {A.}~\bibnamefont {Bur}},
  \bibinfo {author} {\bibfnamefont {T.}~\bibnamefont {Wu}}, \bibinfo {author}
  {\bibfnamefont {N.}~\bibnamefont {Pilet}}, \bibinfo {author} {\bibfnamefont
  {P.}~\bibnamefont {Warnicke}}, \bibinfo {author} {\bibfnamefont {G.~P.}\
  \bibnamefont {Carman}}, \bibinfo {author} {\bibfnamefont {L.~J.}\
  \bibnamefont {Heyderman}}, \ and\ \bibinfo {author} {\bibfnamefont
  {F.}~\bibnamefont {Nolting}},\ }\bibfield  {title} {\enquote {\bibinfo
  {title} {Single domain spin manipulation by electric fields in strain coupled
  artificial multiferroic nanostructures},}\ }\href {\doibase
  https://doi.org/10.1103/PhysRevLett.111.027204} {\bibfield  {journal}
  {\bibinfo  {journal} {Phys. Rev. Lett.}\ }\textbf {\bibinfo {volume} {111}},\
  \bibinfo {pages} {027204} (\bibinfo {year} {2013})}\BibitemShut {NoStop}%
\bibitem [{\citenamefont {Franke}\ \emph {et~al.}(2015)\citenamefont {Franke},
  \citenamefont {Van~de Wiele}, \citenamefont {Shirahata}, \citenamefont
  {H{\"a}m{\"a}l{\"a}inen}, \citenamefont {Taniyama},\ and\ \citenamefont {van
  Dijken}}]{franke-PRX2015}%
  \BibitemOpen
  \bibfield  {author} {\bibinfo {author} {\bibfnamefont {K.~J.~A.}\
  \bibnamefont {Franke}}, \bibinfo {author} {\bibfnamefont {B.}~\bibnamefont
  {Van~de Wiele}}, \bibinfo {author} {\bibfnamefont {Y.}~\bibnamefont
  {Shirahata}}, \bibinfo {author} {\bibfnamefont {S.~J.}\ \bibnamefont
  {H{\"a}m{\"a}l{\"a}inen}}, \bibinfo {author} {\bibfnamefont {T.}~\bibnamefont
  {Taniyama}}, \ and\ \bibinfo {author} {\bibfnamefont {S.}~\bibnamefont {van
  Dijken}},\ }\bibfield  {title} {\enquote {\bibinfo {title} {Reversible
  electric-field-driven magnetic domain-wall motion},}\ }\href {\doibase
  https://doi.org/10.1103/PhysRevX.5.011010} {\bibfield  {journal} {\bibinfo
  {journal} {Phys. Rev. X}\ }\textbf {\bibinfo {volume} {5}},\ \bibinfo {pages}
  {011010} (\bibinfo {year} {2015})}\BibitemShut {NoStop}%
\bibitem [{\citenamefont {LoConte}\ \emph {et~al.}(2018)\citenamefont
  {LoConte}, \citenamefont {Gorchon}, \citenamefont {Mougin}, \citenamefont
  {Lambert}, \citenamefont {El-Ghazaly}, \citenamefont {Scholl}, \citenamefont
  {Salahuddin},\ and\ \citenamefont {Bokor}}]{LoContePRMater:2018}%
  \BibitemOpen
  \bibfield  {author} {\bibinfo {author} {\bibfnamefont {R.}~\bibnamefont
  {LoConte}}, \bibinfo {author} {\bibfnamefont {J.}~\bibnamefont {Gorchon}},
  \bibinfo {author} {\bibfnamefont {A.}~\bibnamefont {Mougin}}, \bibinfo
  {author} {\bibfnamefont {C.~H.~A.}\ \bibnamefont {Lambert}}, \bibinfo
  {author} {\bibfnamefont {A.}~\bibnamefont {El-Ghazaly}}, \bibinfo {author}
  {\bibfnamefont {A.}~\bibnamefont {Scholl}}, \bibinfo {author} {\bibfnamefont
  {S.}~\bibnamefont {Salahuddin}}, \ and\ \bibinfo {author} {\bibfnamefont
  {J.}~\bibnamefont {Bokor}},\ }\bibfield  {title} {\enquote {\bibinfo {title}
  {Electrically controlled switching of the magnetization state in multiferroic
  {B}a{T}i{O}$_3$/{C}o{F}e submicrometer structures},}\ }\href {\doibase
  10.1103/PhysRevMaterials.2.091402} {\bibfield  {journal} {\bibinfo  {journal}
  {Phys. Rev. Mater.}\ }\textbf {\bibinfo {volume} {2}},\ \bibinfo {pages}
  {091402} (\bibinfo {year} {2018})}\BibitemShut {NoStop}%
\bibitem [{\citenamefont {Cavaco}\ \emph {et~al.}(2007)\citenamefont {Cavaco},
  \citenamefont {Van~Kampen}, \citenamefont {Lagae},\ and\ \citenamefont
  {Borghs}}]{Cavaco2007}%
  \BibitemOpen
  \bibfield  {author} {\bibinfo {author} {\bibfnamefont {C.}~\bibnamefont
  {Cavaco}}, \bibinfo {author} {\bibfnamefont {M.}~\bibnamefont {Van~Kampen}},
  \bibinfo {author} {\bibfnamefont {L.}~\bibnamefont {Lagae}}, \ and\ \bibinfo
  {author} {\bibfnamefont {G.}~\bibnamefont {Borghs}},\ }\bibfield  {title}
  {\enquote {\bibinfo {title} {A room-temperature electrical field--controlled
  magnetic memory cell},}\ }\href {\doibase
  https://doi.org/10.1557/jmr.2007.0274} {\bibfield  {journal} {\bibinfo
  {journal} {J. Mater. Res.}\ }\textbf {\bibinfo {volume} {22}},\ \bibinfo
  {pages} {2111--2115} (\bibinfo {year} {2007})}\BibitemShut {NoStop}%
\bibitem [{\citenamefont {Pertsev}\ and\ \citenamefont
  {Kohlstedt}(2009)}]{Pertsev-APL2009}%
  \BibitemOpen
  \bibfield  {author} {\bibinfo {author} {\bibfnamefont {N.~A.}\ \bibnamefont
  {Pertsev}}\ and\ \bibinfo {author} {\bibfnamefont {H.}~\bibnamefont
  {Kohlstedt}},\ }\bibfield  {title} {\enquote {\bibinfo {title} {Magnetic
  tunnel junction on a ferroelectric substrate},}\ }\href {\doibase
  https://doi.org/10.1063/1.3253706} {\bibfield  {journal} {\bibinfo  {journal}
  {Appl. Phys. Lett.}\ }\textbf {\bibinfo {volume} {95}},\ \bibinfo {pages}
  {163503} (\bibinfo {year} {2009})}\BibitemShut {NoStop}%
\bibitem [{\citenamefont {Pertsev}\ and\ \citenamefont
  {Kohlstedt}(2010)}]{Pertsev-Nanotech2010}%
  \BibitemOpen
  \bibfield  {author} {\bibinfo {author} {\bibfnamefont {N.~A.}\ \bibnamefont
  {Pertsev}}\ and\ \bibinfo {author} {\bibfnamefont {H.}~\bibnamefont
  {Kohlstedt}},\ }\bibfield  {title} {\enquote {\bibinfo {title} {Resistive
  switching via the converse magnetoelectric effect in ferromagnetic
  multilayers on ferroelectric substrates},}\ }\href {\doibase
  https://doi.org/10.1088/0957-4484/21/47/475202} {\bibfield  {journal}
  {\bibinfo  {journal} {Nanotechnology}\ }\textbf {\bibinfo {volume} {21}},\
  \bibinfo {pages} {475202} (\bibinfo {year} {2010})}\BibitemShut {NoStop}%
\bibitem [{\citenamefont {Liu}\ \emph {et~al.}(2011)\citenamefont {Liu},
  \citenamefont {Li}, \citenamefont {Obi}, \citenamefont {Lou}, \citenamefont
  {Rand},\ and\ \citenamefont {Sun}}]{LiuAPL:2011}%
  \BibitemOpen
  \bibfield  {author} {\bibinfo {author} {\bibfnamefont {M.}~\bibnamefont
  {Liu}}, \bibinfo {author} {\bibfnamefont {S.}~\bibnamefont {Li}}, \bibinfo
  {author} {\bibfnamefont {O.}~\bibnamefont {Obi}}, \bibinfo {author}
  {\bibfnamefont {J.}~\bibnamefont {Lou}}, \bibinfo {author} {\bibfnamefont
  {S.}~\bibnamefont {Rand}}, \ and\ \bibinfo {author} {\bibfnamefont {N.~X.}\
  \bibnamefont {Sun}},\ }\bibfield  {title} {\enquote {\bibinfo {title}
  {Electric field modulation of magnetoresistance in multiferroic
  heterostructures for ultralow power electronics},}\ }\href {\doibase
  https://doi.org/10.1063/1.3597796} {\bibfield  {journal} {\bibinfo  {journal}
  {Appl. Phys. Lett.}\ }\textbf {\bibinfo {volume} {98}},\ \bibinfo {pages}
  {222509} (\bibinfo {year} {2011})}\BibitemShut {NoStop}%
\bibitem [{\citenamefont {Chen}\ \emph {et~al.}(2019)\citenamefont {Chen},
  \citenamefont {Wen}, \citenamefont {Fang}, \citenamefont {Zhao},
  \citenamefont {Zhang}, \citenamefont {Chang}, \citenamefont {Li},
  \citenamefont {Wu}, \citenamefont {Huang}, \citenamefont {Lu}, \citenamefont
  {Zeng}, \citenamefont {Cai}, \citenamefont {Han}, \citenamefont {Wu},
  \citenamefont {Zhang},\ and\ \citenamefont {Zhao}}]{Chen-NComm2019}%
  \BibitemOpen
  \bibfield  {author} {\bibinfo {author} {\bibfnamefont {A.}~\bibnamefont
  {Chen}}, \bibinfo {author} {\bibfnamefont {Y.}~\bibnamefont {Wen}}, \bibinfo
  {author} {\bibfnamefont {B.}~\bibnamefont {Fang}}, \bibinfo {author}
  {\bibfnamefont {Y.}~\bibnamefont {Zhao}}, \bibinfo {author} {\bibfnamefont
  {Q.}~\bibnamefont {Zhang}}, \bibinfo {author} {\bibfnamefont
  {Y.}~\bibnamefont {Chang}}, \bibinfo {author} {\bibfnamefont
  {P.}~\bibnamefont {Li}}, \bibinfo {author} {\bibfnamefont {H.}~\bibnamefont
  {Wu}}, \bibinfo {author} {\bibfnamefont {H.}~\bibnamefont {Huang}}, \bibinfo
  {author} {\bibfnamefont {Y.}~\bibnamefont {Lu}}, \bibinfo {author}
  {\bibfnamefont {Z.}~\bibnamefont {Zeng}}, \bibinfo {author} {\bibfnamefont
  {J.}~\bibnamefont {Cai}}, \bibinfo {author} {\bibfnamefont {X.}~\bibnamefont
  {Han}}, \bibinfo {author} {\bibfnamefont {T.}~\bibnamefont {Wu}}, \bibinfo
  {author} {\bibfnamefont {X.-X.}\ \bibnamefont {Zhang}}, \ and\ \bibinfo
  {author} {\bibfnamefont {Y.}~\bibnamefont {Zhao}},\ }\bibfield  {title}
  {\enquote {\bibinfo {title} {Giant nonvolatile manipulation of
  magnetoresistance in magnetic tunnel junctions by electric fields via
  magnetoelectric coupling},}\ }\href {\doibase
  https://doi.org/10.1038/s41467-018-08061-5} {\bibfield  {journal} {\bibinfo
  {journal} {Nat. Commun.}\ }\textbf {\bibinfo {volume} {10}},\ \bibinfo
  {pages} {1--7} (\bibinfo {year} {2019})}\BibitemShut {NoStop}%
\bibitem [{\citenamefont {Liu}\ \emph {et~al.}(2013)\citenamefont {Liu},
  \citenamefont {Zhou}, \citenamefont {Nan}, \citenamefont {Howe},
  \citenamefont {Brown},\ and\ \citenamefont {Sun}}]{liu-AdvMat2013}%
  \BibitemOpen
  \bibfield  {author} {\bibinfo {author} {\bibfnamefont {M.}~\bibnamefont
  {Liu}}, \bibinfo {author} {\bibfnamefont {Z.}~\bibnamefont {Zhou}}, \bibinfo
  {author} {\bibfnamefont {T.}~\bibnamefont {Nan}}, \bibinfo {author}
  {\bibfnamefont {B.~M.}\ \bibnamefont {Howe}}, \bibinfo {author}
  {\bibfnamefont {G.~J.}\ \bibnamefont {Brown}}, \ and\ \bibinfo {author}
  {\bibfnamefont {N.~X.}\ \bibnamefont {Sun}},\ }\bibfield  {title} {\enquote
  {\bibinfo {title} {Voltage tuning of ferromagnetic resonance with bistable
  magnetization switching in energy-efficient magnetoelectric composites},}\
  }\href {\doibase https://doi.org/10.1002/adma.201203792} {\bibfield
  {journal} {\bibinfo  {journal} {Adv. Mater.}\ }\textbf {\bibinfo {volume}
  {25}},\ \bibinfo {pages} {1435--1439} (\bibinfo {year} {2013})}\BibitemShut
  {NoStop}%
\bibitem [{\citenamefont {Lou}\ \emph {et~al.}(2009)\citenamefont {Lou},
  \citenamefont {Liu}, \citenamefont {Reed}, \citenamefont {Ren},\ and\
  \citenamefont {Sun}}]{LouAdMat:2009}%
  \BibitemOpen
  \bibfield  {author} {\bibinfo {author} {\bibfnamefont {J.}~\bibnamefont
  {Lou}}, \bibinfo {author} {\bibfnamefont {M.}~\bibnamefont {Liu}}, \bibinfo
  {author} {\bibfnamefont {D.}~\bibnamefont {Reed}}, \bibinfo {author}
  {\bibfnamefont {Y.}~\bibnamefont {Ren}}, \ and\ \bibinfo {author}
  {\bibfnamefont {N.~X.}\ \bibnamefont {Sun}},\ }\bibfield  {title} {\enquote
  {\bibinfo {title} {Giant electric field tuning of magnetism in novel
  multiferroic {F}e{G}a{B}/lead zinc niobate–lead titanate ({PZN}-{PT})
  heterostructures},}\ }\href {\doibase 10.1002/adma.200901131} {\bibfield
  {journal} {\bibinfo  {journal} {Adv. Mater.}\ }\textbf {\bibinfo {volume}
  {21}},\ \bibinfo {pages} {4711--4715} (\bibinfo {year} {2009})}\BibitemShut
  {NoStop}%
\bibitem [{\citenamefont {Brandl}\ \emph {et~al.}(2014)\citenamefont {Brandl},
  \citenamefont {Franke}, \citenamefont {Lahtinen}, \citenamefont {van
  Dijken},\ and\ \citenamefont {Grundler}}]{Brandl:2014}%
  \BibitemOpen
  \bibfield  {author} {\bibinfo {author} {\bibfnamefont {F.}~\bibnamefont
  {Brandl}}, \bibinfo {author} {\bibfnamefont {K.~J.~A.}\ \bibnamefont
  {Franke}}, \bibinfo {author} {\bibfnamefont {T.~H.~E.}\ \bibnamefont
  {Lahtinen}}, \bibinfo {author} {\bibfnamefont {S.}~\bibnamefont {van
  Dijken}}, \ and\ \bibinfo {author} {\bibfnamefont {D.}~\bibnamefont
  {Grundler}},\ }\bibfield  {title} {\enquote {\bibinfo {title} {Spin waves in
  {C}o{F}e{B} on ferroelectric domains combining spin mechanics and
  magnonics},}\ }\href {\doibase https://doi.org/10.1016/j.ssc.2013.12.019}
  {\bibfield  {journal} {\bibinfo  {journal} {Solid State Commun.}\ }\textbf
  {\bibinfo {volume} {198}},\ \bibinfo {pages} {13--17} (\bibinfo {year}
  {2014})}\BibitemShut {NoStop}%
\bibitem [{\citenamefont {Azovtsev}\ and\ \citenamefont
  {Pertsev}(2018)}]{Azovtsev-PRAppl2018}%
  \BibitemOpen
  \bibfield  {author} {\bibinfo {author} {\bibfnamefont {A.~V.}\ \bibnamefont
  {Azovtsev}}\ and\ \bibinfo {author} {\bibfnamefont {N.~A.}\ \bibnamefont
  {Pertsev}},\ }\bibfield  {title} {\enquote {\bibinfo {title} {Electrical
  tuning of ferromagnetic resonance in thin-film nanomagnets coupled to
  piezoelectrically active substrates},}\ }\href {\doibase
  10.1103/PhysRevApplied.10.044041} {\bibfield  {journal} {\bibinfo  {journal}
  {Phys. Rev. Appl.}\ }\textbf {\bibinfo {volume} {10}},\ \bibinfo {pages}
  {044041} (\bibinfo {year} {2018})}\BibitemShut {NoStop}%
\bibitem [{\citenamefont {H{\"a}m{\"a}l{\"a}inen}\ \emph
  {et~al.}(2018)\citenamefont {H{\"a}m{\"a}l{\"a}inen}, \citenamefont {Madami},
  \citenamefont {Qin}, \citenamefont {Gubbiotti},\ and\ \citenamefont {van
  Dijken}}]{HamalainenNature:2018}%
  \BibitemOpen
  \bibfield  {author} {\bibinfo {author} {\bibfnamefont {S.~J}\ \bibnamefont
  {H{\"a}m{\"a}l{\"a}inen}}, \bibinfo {author} {\bibfnamefont {M.}~\bibnamefont
  {Madami}}, \bibinfo {author} {\bibfnamefont {H.}~\bibnamefont {Qin}},
  \bibinfo {author} {\bibfnamefont {G.}~\bibnamefont {Gubbiotti}}, \ and\
  \bibinfo {author} {\bibfnamefont {S.}~\bibnamefont {van Dijken}},\ }\bibfield
   {title} {\enquote {\bibinfo {title} {Control of spin-wave transmission by a
  programmable domain wall},}\ }\href {\doibase
  doi.org/10.1038/s41467-018-07372-x} {\bibfield  {journal} {\bibinfo
  {journal} {Nat. Commun.}\ }\textbf {\bibinfo {volume} {9}},\ \bibinfo {pages}
  {1--8} (\bibinfo {year} {2018})}\BibitemShut {NoStop}%
\bibitem [{\citenamefont {Sadovnikov}\ \emph {et~al.}(2018)\citenamefont
  {Sadovnikov}, \citenamefont {Grachev}, \citenamefont {Sheshukova},
  \citenamefont {Sharaevskii}, \citenamefont {Serdobintsev}, \citenamefont
  {Mitin},\ and\ \citenamefont {Nikitov}}]{sadovnikov-PRL2018}%
  \BibitemOpen
  \bibfield  {author} {\bibinfo {author} {\bibfnamefont {A.~V.}\ \bibnamefont
  {Sadovnikov}}, \bibinfo {author} {\bibfnamefont {A.~A.}\ \bibnamefont
  {Grachev}}, \bibinfo {author} {\bibfnamefont {S.~E.}\ \bibnamefont
  {Sheshukova}}, \bibinfo {author} {\bibfnamefont {Yu.~P.}\ \bibnamefont
  {Sharaevskii}}, \bibinfo {author} {\bibfnamefont {A.~A.}\ \bibnamefont
  {Serdobintsev}}, \bibinfo {author} {\bibfnamefont {D.~M.}\ \bibnamefont
  {Mitin}}, \ and\ \bibinfo {author} {\bibfnamefont {S.~A.}\ \bibnamefont
  {Nikitov}},\ }\bibfield  {title} {\enquote {\bibinfo {title} {Magnon
  straintronics: reconfigurable spin-wave routing in strain-controlled
  bilateral magnetic stripes},}\ }\href {\doibase
  https://doi.org/10.1103/PhysRevLett.120.257203} {\bibfield  {journal}
  {\bibinfo  {journal} {Phys. Rev. Lett.}\ }\textbf {\bibinfo {volume} {120}},\
  \bibinfo {pages} {257203} (\bibinfo {year} {2018})}\BibitemShut {NoStop}%
\bibitem [{\citenamefont {Savostin}\ and\ \citenamefont
  {Pertsev}(2020)}]{Savostin-Nanoscale2020}%
  \BibitemOpen
  \bibfield  {author} {\bibinfo {author} {\bibfnamefont {E.~O.}\ \bibnamefont
  {Savostin}}\ and\ \bibinfo {author} {\bibfnamefont {N.~A.}\ \bibnamefont
  {Pertsev}},\ }\bibfield  {title} {\enquote {\bibinfo {title} {Superconducting
  straintronics via the proximity effect in superconductor--ferromagnet
  nanostructures},}\ }\href {\doibase 10.1039/C9NR06739F} {\bibfield  {journal}
  {\bibinfo  {journal} {Nanoscale}\ }\textbf {\bibinfo {volume} {12}},\
  \bibinfo {pages} {648--657} (\bibinfo {year} {2020})}\BibitemShut {NoStop}%
\bibitem [{\citenamefont {Kirilyuk}\ \emph {et~al.}(2010)\citenamefont
  {Kirilyuk}, \citenamefont {Kimel},\ and\ \citenamefont
  {Rasing}}]{Kirilyuk-RMP2010}%
  \BibitemOpen
  \bibfield  {author} {\bibinfo {author} {\bibfnamefont {A.}~\bibnamefont
  {Kirilyuk}}, \bibinfo {author} {\bibfnamefont {A.~V.}\ \bibnamefont {Kimel}},
  \ and\ \bibinfo {author} {\bibfnamefont {Th.}\ \bibnamefont {Rasing}},\
  }\bibfield  {title} {\enquote {\bibinfo {title} {Ultrafast optical
  manipulation of magnetic order},}\ }\href {\doibase
  10.1103/RevModPhys.82.2731} {\bibfield  {journal} {\bibinfo  {journal} {Rev.
  Mod. Phys.}\ }\textbf {\bibinfo {volume} {82}},\ \bibinfo {pages}
  {2731--2784} (\bibinfo {year} {2010})}\BibitemShut {NoStop}%
\bibitem [{\citenamefont {Sheu}\ \emph {et~al.}(2014)\citenamefont {Sheu},
  \citenamefont {Trugman}, \citenamefont {Yan}, \citenamefont {Jia},
  \citenamefont {Taylor},\ and\ \citenamefont {Prasankumar}}]{Sheu:2014}%
  \BibitemOpen
  \bibfield  {author} {\bibinfo {author} {\bibfnamefont {Y.~M.}\ \bibnamefont
  {Sheu}}, \bibinfo {author} {\bibfnamefont {S.~A.}\ \bibnamefont {Trugman}},
  \bibinfo {author} {\bibfnamefont {L.}~\bibnamefont {Yan}}, \bibinfo {author}
  {\bibfnamefont {Q.~X.}\ \bibnamefont {Jia}}, \bibinfo {author} {\bibfnamefont
  {A.~J.}\ \bibnamefont {Taylor}}, \ and\ \bibinfo {author} {\bibfnamefont
  {R.~P.}\ \bibnamefont {Prasankumar}},\ }\bibfield  {title} {\enquote
  {\bibinfo {title} {Using ultrashort optical pulses to couple ferroelectric
  and ferromagnetic order in an oxide heterostructure},}\ }\href {\doibase
  https://doi.org/10.1038/ncomms6832} {\bibfield  {journal} {\bibinfo
  {journal} {Nat. Commun.}\ }\textbf {\bibinfo {volume} {5}},\ \bibinfo {pages}
  {1--6} (\bibinfo {year} {2014})}\BibitemShut {NoStop}%
\bibitem [{\citenamefont {Jia}\ \emph {et~al.}(2016)\citenamefont {Jia},
  \citenamefont {Zhang}, \citenamefont {Sukhov},\ and\ \citenamefont
  {Berakdar}}]{Jia:2016}%
  \BibitemOpen
  \bibfield  {author} {\bibinfo {author} {\bibfnamefont {C.}~\bibnamefont
  {Jia}}, \bibinfo {author} {\bibfnamefont {N.}~\bibnamefont {Zhang}}, \bibinfo
  {author} {\bibfnamefont {A.}~\bibnamefont {Sukhov}}, \ and\ \bibinfo {author}
  {\bibfnamefont {J.}~\bibnamefont {Berakdar}},\ }\bibfield  {title} {\enquote
  {\bibinfo {title} {Ultrafast transient dynamics in composite
  multiferroics},}\ }\href {\doibase
  https://doi.org/10.1088/1367-2630/18/2/023002} {\bibfield  {journal}
  {\bibinfo  {journal} {New J. Phys.}\ }\textbf {\bibinfo {volume} {18}},\
  \bibinfo {pages} {023002} (\bibinfo {year} {2016})}\BibitemShut {NoStop}%
\bibitem [{\citenamefont {Lejman}\ \emph {et~al.}(2014)\citenamefont {Lejman},
  \citenamefont {Vaudel}, \citenamefont {Infante}, \citenamefont {Gemeiner},
  \citenamefont {Gusev}, \citenamefont {Dkhil},\ and\ \citenamefont
  {Ruello}}]{Lejman-NatureComm2014}%
  \BibitemOpen
  \bibfield  {author} {\bibinfo {author} {\bibfnamefont {M.}~\bibnamefont
  {Lejman}}, \bibinfo {author} {\bibfnamefont {G.}~\bibnamefont {Vaudel}},
  \bibinfo {author} {\bibfnamefont {I.~C.}\ \bibnamefont {Infante}}, \bibinfo
  {author} {\bibfnamefont {P.}~\bibnamefont {Gemeiner}}, \bibinfo {author}
  {\bibfnamefont {V.~E.}\ \bibnamefont {Gusev}}, \bibinfo {author}
  {\bibfnamefont {B.}~\bibnamefont {Dkhil}}, \ and\ \bibinfo {author}
  {\bibfnamefont {P.}~\bibnamefont {Ruello}},\ }\bibfield  {title} {\enquote
  {\bibinfo {title} {Giant ultrafast photo-induced shear strain in
  ferroelectric {B}i{F}e{O}$_3$},}\ }\href {\doibase
  https://doi.org/10.1038/ncomms5301} {\bibfield  {journal} {\bibinfo
  {journal} {Nat. Commun.}\ }\textbf {\bibinfo {volume} {5}},\ \bibinfo {pages}
  {1--7} (\bibinfo {year} {2014})}\BibitemShut {NoStop}%
\bibitem [{\citenamefont {Kimel}\ \emph {et~al.}(2020)\citenamefont {Kimel},
  \citenamefont {Kalashnikova}, \citenamefont {Pogrebna},\ and\ \citenamefont
  {Zvezdin}}]{KIMEL2020}%
  \BibitemOpen
  \bibfield  {author} {\bibinfo {author} {\bibfnamefont {A.~V.}\ \bibnamefont
  {Kimel}}, \bibinfo {author} {\bibfnamefont {A.~M.}\ \bibnamefont
  {Kalashnikova}}, \bibinfo {author} {\bibfnamefont {A.}~\bibnamefont
  {Pogrebna}}, \ and\ \bibinfo {author} {\bibfnamefont {A.~K.}\ \bibnamefont
  {Zvezdin}},\ }\bibfield  {title} {\enquote {\bibinfo {title} {Fundamentals
  and perspectives of ultrafast photoferroic recording},}\ }\href {\doibase
  https://doi.org/10.1016/j.physrep.2020.01.004} {\bibfield  {journal}
  {\bibinfo  {journal} {Phys. Rep.}\ }\textbf {\bibinfo {volume} {852}},\
  \bibinfo {pages} {1--46} (\bibinfo {year} {2020})}\BibitemShut {NoStop}%
\bibitem [{\citenamefont {Liu}\ \emph {et~al.}(2012)\citenamefont {Liu},
  \citenamefont {Chen}, \citenamefont {He}, \citenamefont {Liang},
  \citenamefont {Chen}, \citenamefont {Chien}, \citenamefont {Hsieh},
  \citenamefont {Lin}, \citenamefont {Arenholz}, \citenamefont {Luo},
  \citenamefont {Chueh}, \citenamefont {Chen},\ and\ \citenamefont
  {Chu}}]{Liu-ACSNano2012}%
  \BibitemOpen
  \bibfield  {author} {\bibinfo {author} {\bibfnamefont {H.-J.}\ \bibnamefont
  {Liu}}, \bibinfo {author} {\bibfnamefont {L.-Y.}\ \bibnamefont {Chen}},
  \bibinfo {author} {\bibfnamefont {Q.}~\bibnamefont {He}}, \bibinfo {author}
  {\bibfnamefont {C.-W.}\ \bibnamefont {Liang}}, \bibinfo {author}
  {\bibfnamefont {Y.-Z.}\ \bibnamefont {Chen}}, \bibinfo {author}
  {\bibfnamefont {Y.-S.}\ \bibnamefont {Chien}}, \bibinfo {author}
  {\bibfnamefont {Y.-H.}\ \bibnamefont {Hsieh}}, \bibinfo {author}
  {\bibfnamefont {S.-J.}\ \bibnamefont {Lin}}, \bibinfo {author} {\bibfnamefont
  {E.}~\bibnamefont {Arenholz}}, \bibinfo {author} {\bibfnamefont {C.-W.}\
  \bibnamefont {Luo}}, \bibinfo {author} {\bibfnamefont {Y.-L.}\ \bibnamefont
  {Chueh}}, \bibinfo {author} {\bibfnamefont {Y.-C.}\ \bibnamefont {Chen}}, \
  and\ \bibinfo {author} {\bibfnamefont {Y.-H.}\ \bibnamefont {Chu}},\
  }\bibfield  {title} {\enquote {\bibinfo {title} {Epitaxial
  photostriction–magnetostriction coupled self-assembled nanostructures},}\
  }\href {\doibase 10.1021/nn301976p} {\bibfield  {journal} {\bibinfo
  {journal} {ACS Nano}\ }\textbf {\bibinfo {volume} {6}},\ \bibinfo {pages}
  {6952--6959} (\bibinfo {year} {2012})}\BibitemShut {NoStop}%
\bibitem [{\citenamefont {Bigot}\ \emph {et~al.}(2005)\citenamefont {Bigot},
  \citenamefont {Vomir}, \citenamefont {Andrade},\ and\ \citenamefont
  {Beaurepaire}}]{Bigot-ChemPhys2005}%
  \BibitemOpen
  \bibfield  {author} {\bibinfo {author} {\bibfnamefont {J.-Y.}\ \bibnamefont
  {Bigot}}, \bibinfo {author} {\bibfnamefont {M.}~\bibnamefont {Vomir}},
  \bibinfo {author} {\bibfnamefont {L.~H.~F.}\ \bibnamefont {Andrade}}, \ and\
  \bibinfo {author} {\bibfnamefont {E.}~\bibnamefont {Beaurepaire}},\
  }\bibfield  {title} {\enquote {\bibinfo {title} {Ultrafast magnetization
  dynamics in ferromagnetic cobalt: The role of the anisotropy},}\ }\href
  {\doibase https://doi.org/10.1016/j.chemphys.2005.06.016} {\bibfield
  {journal} {\bibinfo  {journal} {Chem. Phys.}\ }\textbf {\bibinfo {volume}
  {318}},\ \bibinfo {pages} {137--146} (\bibinfo {year} {2005})}\BibitemShut
  {NoStop}%
\bibitem [{\citenamefont {Carpene}\ \emph
  {et~al.}(2010{\natexlab{a}})\citenamefont {Carpene}, \citenamefont {Mancini},
  \citenamefont {Dazzi}, \citenamefont {Dallera}, \citenamefont {Puppin},\ and\
  \citenamefont {De~Silvestri}}]{CarpenePRB:2010}%
  \BibitemOpen
  \bibfield  {author} {\bibinfo {author} {\bibfnamefont {E.}~\bibnamefont
  {Carpene}}, \bibinfo {author} {\bibfnamefont {E.}~\bibnamefont {Mancini}},
  \bibinfo {author} {\bibfnamefont {D.}~\bibnamefont {Dazzi}}, \bibinfo
  {author} {\bibfnamefont {C.}~\bibnamefont {Dallera}}, \bibinfo {author}
  {\bibfnamefont {E.}~\bibnamefont {Puppin}}, \ and\ \bibinfo {author}
  {\bibfnamefont {S.}~\bibnamefont {De~Silvestri}},\ }\bibfield  {title}
  {\enquote {\bibinfo {title} {Ultrafast three-dimensional magnetization
  precession and magnetic anisotropy of a photoexcited thin film of iron},}\
  }\href {\doibase 10.1103/PhysRevB.81.060415} {\bibfield  {journal} {\bibinfo
  {journal} {Phys. Rev. B}\ }\textbf {\bibinfo {volume} {81}},\ \bibinfo
  {pages} {060415(R)} (\bibinfo {year} {2010}{\natexlab{a}})}\BibitemShut
  {NoStop}%
\bibitem [{\citenamefont {Carpene}\ \emph
  {et~al.}(2010{\natexlab{b}})\citenamefont {Carpene}, \citenamefont {Mancini},
  \citenamefont {Dallera}, \citenamefont {Puppin},\ and\ \citenamefont
  {De~Silvestri}}]{CarpeneJAP:2010}%
  \BibitemOpen
  \bibfield  {author} {\bibinfo {author} {\bibfnamefont {E.}~\bibnamefont
  {Carpene}}, \bibinfo {author} {\bibfnamefont {E.}~\bibnamefont {Mancini}},
  \bibinfo {author} {\bibfnamefont {C.}~\bibnamefont {Dallera}}, \bibinfo
  {author} {\bibfnamefont {E.}~\bibnamefont {Puppin}}, \ and\ \bibinfo {author}
  {\bibfnamefont {S.}~\bibnamefont {De~Silvestri}},\ }\bibfield  {title}
  {\enquote {\bibinfo {title} {Three-dimensional magnetization evolution and
  the role of anisotropies in thin {F}e/{M}g{O} films: Static and dynamic
  measurements},}\ }\href {\doibase 10.1063/1.3488639} {\bibfield  {journal}
  {\bibinfo  {journal} {J. Appl. Phys.}\ }\textbf {\bibinfo {volume} {108}},\
  \bibinfo {pages} {063919} (\bibinfo {year} {2010}{\natexlab{b}})}\BibitemShut
  {NoStop}%
\bibitem [{\citenamefont {Shelukhin}\ \emph {et~al.}(2018)\citenamefont
  {Shelukhin}, \citenamefont {Pavlov}, \citenamefont {Usachev}, \citenamefont
  {Shamray}, \citenamefont {Pisarev},\ and\ \citenamefont
  {Kalashnikova}}]{Shelukhin:PRB2018}%
  \BibitemOpen
  \bibfield  {author} {\bibinfo {author} {\bibfnamefont {L.~A.}\ \bibnamefont
  {Shelukhin}}, \bibinfo {author} {\bibfnamefont {V.~V.}\ \bibnamefont
  {Pavlov}}, \bibinfo {author} {\bibfnamefont {P.~A.}\ \bibnamefont {Usachev}},
  \bibinfo {author} {\bibfnamefont {P.~Yu.}\ \bibnamefont {Shamray}}, \bibinfo
  {author} {\bibfnamefont {R.~V.}\ \bibnamefont {Pisarev}}, \ and\ \bibinfo
  {author} {\bibfnamefont {A.~M.}\ \bibnamefont {Kalashnikova}},\ }\bibfield
  {title} {\enquote {\bibinfo {title} {Ultrafast laser-induced changes of the
  magnetic anisotropy in a low-symmetry iron garnet film},}\ }\href {\doibase
  10.1103/PhysRevB.97.014422} {\bibfield  {journal} {\bibinfo  {journal} {Phys.
  Rev.B}\ }\textbf {\bibinfo {volume} {97}},\ \bibinfo {pages} {014422}
  (\bibinfo {year} {2018})}\BibitemShut {NoStop}%
\bibitem [{\citenamefont {Pertsev}(2008)}]{Pertsev-PRB2008}%
  \BibitemOpen
  \bibfield  {author} {\bibinfo {author} {\bibfnamefont {N.~A.}\ \bibnamefont
  {Pertsev}},\ }\bibfield  {title} {\enquote {\bibinfo {title} {Giant
  magnetoelectric effect via strain-induced spin reorientation transitions in
  ferromagnetic films},}\ }\href {\doibase 10.1103/PhysRevB.78.212102}
  {\bibfield  {journal} {\bibinfo  {journal} {Phys. Rev. B}\ }\textbf {\bibinfo
  {volume} {78}},\ \bibinfo {pages} {212102} (\bibinfo {year}
  {2008})}\BibitemShut {NoStop}%
\bibitem [{\citenamefont {Hirth}\ and\ \citenamefont
  {Lothe}(1968)}]{CoFe-elastic}%
  \BibitemOpen
  \bibfield  {author} {\bibinfo {author} {\bibfnamefont {J.~P.}\ \bibnamefont
  {Hirth}}\ and\ \bibinfo {author} {\bibfnamefont {J.}~\bibnamefont {Lothe}},\
  }\href@noop {} {\emph {\bibinfo {title} {Theory of Dislocations}}}\ (\bibinfo
   {publisher} {McGraw-Hill, New York},\ \bibinfo {year} {1968})\BibitemShut
  {NoStop}%
\bibitem [{\citenamefont {Touloukian}\ \emph {et~al.}(1970)\citenamefont
  {Touloukian}, \citenamefont {Powell}, \citenamefont {Ho},\ and\ \citenamefont
  {Klemens}}]{Touloukian}%
  \BibitemOpen
  \bibfield  {author} {\bibinfo {author} {\bibfnamefont {Y.~S.}\ \bibnamefont
  {Touloukian}}, \bibinfo {author} {\bibfnamefont {R.~W.}\ \bibnamefont
  {Powell}}, \bibinfo {author} {\bibfnamefont {C.~Y.}\ \bibnamefont {Ho}}, \
  and\ \bibinfo {author} {\bibfnamefont {P.~G.}\ \bibnamefont {Klemens}},\
  }\bibfield  {title} {\enquote {\bibinfo {title} {Thermophysical properties of
  matter-the tprc data series. volume 2. thermal conductivity-nonmetallic
  solids},}\ }\href@noop {} {\bibfield  {journal} {\bibinfo  {journal}
  {IFI/Plenum, New York}\ } (\bibinfo {year} {1970})}\BibitemShut {NoStop}%
\bibitem [{\citenamefont {An}\ \emph {et~al.}(2016)\citenamefont {An},
  \citenamefont {Ma}, \citenamefont {Pai}, \citenamefont {Yang}, \citenamefont
  {Olsson}, \citenamefont {Erskine}, \citenamefont {Ralph}, \citenamefont
  {Buhrman},\ and\ \citenamefont {Li}}]{An-PRB2016}%
  \BibitemOpen
  \bibfield  {author} {\bibinfo {author} {\bibfnamefont {K.}~\bibnamefont
  {An}}, \bibinfo {author} {\bibfnamefont {X.}~\bibnamefont {Ma}}, \bibinfo
  {author} {\bibfnamefont {C.-F.}\ \bibnamefont {Pai}}, \bibinfo {author}
  {\bibfnamefont {J.}~\bibnamefont {Yang}}, \bibinfo {author} {\bibfnamefont
  {K.~S.}\ \bibnamefont {Olsson}}, \bibinfo {author} {\bibfnamefont {J.~L.}\
  \bibnamefont {Erskine}}, \bibinfo {author} {\bibfnamefont {D.~C.}\
  \bibnamefont {Ralph}}, \bibinfo {author} {\bibfnamefont {R.~A.}\ \bibnamefont
  {Buhrman}}, \ and\ \bibinfo {author} {\bibfnamefont {X.}~\bibnamefont {Li}},\
  }\bibfield  {title} {\enquote {\bibinfo {title} {Current control of magnetic
  anisotropy via stress in a ferromagnetic metal waveguide},}\ }\href {\doibase
  10.1103/PhysRevB.93.140404} {\bibfield  {journal} {\bibinfo  {journal} {Phys.
  Rev. B}\ }\textbf {\bibinfo {volume} {93}},\ \bibinfo {pages} {140404(R)}
  (\bibinfo {year} {2016})}\BibitemShut {NoStop}%
\bibitem [{BaT()}]{BaTiO3-crystal}%
  \BibitemOpen
  \href@noop {} {\enquote {\bibinfo {title} {{B}a{T}i{O}$_3$ crystal structure:
  Datasheet from ``{PAULING FILE} {M}ultinaries {E}dition -- 2012'' in
  {S}pringer{M}aterials},}\ }\BibitemShut {NoStop}%
\bibitem [{\citenamefont {Kay}\ and\ \citenamefont {Vousden}(1949)}]{Kay-1949}%
  \BibitemOpen
  \bibfield  {author} {\bibinfo {author} {\bibfnamefont {H.~F.}\ \bibnamefont
  {Kay}}\ and\ \bibinfo {author} {\bibfnamefont {P.}~\bibnamefont {Vousden}},\
  }\bibfield  {title} {\enquote {\bibinfo {title} {Xcv. symmetry changes in
  barium titanate at low temperatures and their relation to its ferroelectric
  properties},}\ }\href {\doibase 10.1080/14786444908561371} {\bibfield
  {journal} {\bibinfo  {journal} {Philos. Mag.}\ }\textbf {\bibinfo {volume}
  {40}},\ \bibinfo {pages} {1019--1040} (\bibinfo {year} {1949})}\BibitemShut
  {NoStop}%
\bibitem [{\citenamefont {Beaurepaire}\ \emph {et~al.}(1996)\citenamefont
  {Beaurepaire}, \citenamefont {Merle}, \citenamefont {Daunois},\ and\
  \citenamefont {Bigot}}]{Beaurepaire:PRL1996}%
  \BibitemOpen
  \bibfield  {author} {\bibinfo {author} {\bibfnamefont {E.}~\bibnamefont
  {Beaurepaire}}, \bibinfo {author} {\bibfnamefont {J.-C.}\ \bibnamefont
  {Merle}}, \bibinfo {author} {\bibfnamefont {A.}~\bibnamefont {Daunois}}, \
  and\ \bibinfo {author} {\bibfnamefont {J.-Y.}\ \bibnamefont {Bigot}},\
  }\bibfield  {title} {\enquote {\bibinfo {title} {Ultrafast spin dynamics in
  ferromagnetic nickel},}\ }\href {\doibase 10.1103/PhysRevLett.76.4250}
  {\bibfield  {journal} {\bibinfo  {journal} {Phys. Rev. Lett.}\ }\textbf
  {\bibinfo {volume} {76}},\ \bibinfo {pages} {4250--4253} (\bibinfo {year}
  {1996})}\BibitemShut {NoStop}%
\bibitem [{\citenamefont {Koopmans}\ \emph {et~al.}(2010)\citenamefont
  {Koopmans}, \citenamefont {Malinowski}, \citenamefont {Dalla~Longa},
  \citenamefont {Steiauf}, \citenamefont {F{\"a}hnle}, \citenamefont {Roth},
  \citenamefont {Cinchetti},\ and\ \citenamefont
  {Aeschlimann}}]{Koopmans:NatureMater2009}%
  \BibitemOpen
  \bibfield  {author} {\bibinfo {author} {\bibfnamefont {B.}~\bibnamefont
  {Koopmans}}, \bibinfo {author} {\bibfnamefont {G.}~\bibnamefont
  {Malinowski}}, \bibinfo {author} {\bibfnamefont {F.}~\bibnamefont
  {Dalla~Longa}}, \bibinfo {author} {\bibfnamefont {D.}~\bibnamefont
  {Steiauf}}, \bibinfo {author} {\bibfnamefont {M.}~\bibnamefont {F{\"a}hnle}},
  \bibinfo {author} {\bibfnamefont {T.}~\bibnamefont {Roth}}, \bibinfo {author}
  {\bibfnamefont {M.}~\bibnamefont {Cinchetti}}, \ and\ \bibinfo {author}
  {\bibfnamefont {M.}~\bibnamefont {Aeschlimann}},\ }\bibfield  {title}
  {\enquote {\bibinfo {title} {Explaining the paradoxical diversity of
  ultrafast laser-induced demagnetization},}\ }\href {\doibase
  10.1038/nmat2593} {\bibfield  {journal} {\bibinfo  {journal} {Nat. Mater.}\
  }\textbf {\bibinfo {volume} {9}},\ \bibinfo {pages} {259--265} (\bibinfo
  {year} {2010})}\BibitemShut {NoStop}%
\bibitem [{\citenamefont {van Kampen}\ \emph {et~al.}(2002)\citenamefont {van
  Kampen}, \citenamefont {Jozsa}, \citenamefont {Kohlhepp}, \citenamefont
  {LeClair}, \citenamefont {Lagae}, \citenamefont {de~Jonge},\ and\
  \citenamefont {Koopmans}}]{vanKampen:PLR2002}%
  \BibitemOpen
  \bibfield  {author} {\bibinfo {author} {\bibfnamefont {M.}~\bibnamefont {van
  Kampen}}, \bibinfo {author} {\bibfnamefont {C.}~\bibnamefont {Jozsa}},
  \bibinfo {author} {\bibfnamefont {J.~T.}\ \bibnamefont {Kohlhepp}}, \bibinfo
  {author} {\bibfnamefont {P.}~\bibnamefont {LeClair}}, \bibinfo {author}
  {\bibfnamefont {L.}~\bibnamefont {Lagae}}, \bibinfo {author} {\bibfnamefont
  {W.~J.~M.}\ \bibnamefont {de~Jonge}}, \ and\ \bibinfo {author} {\bibfnamefont
  {B.}~\bibnamefont {Koopmans}},\ }\bibfield  {title} {\enquote {\bibinfo
  {title} {All-optical probe of coherent spin waves},}\ }\href {\doibase
  10.1103/PhysRevLett.88.227201} {\bibfield  {journal} {\bibinfo  {journal}
  {Phys. Rev. Lett.}\ }\textbf {\bibinfo {volume} {88}},\ \bibinfo {pages}
  {227201} (\bibinfo {year} {2002})}\BibitemShut {NoStop}%
\bibitem [{\citenamefont {Gurevich}\ and\ \citenamefont
  {Melkov}(1996)}]{gurevich1996magnetization}%
  \BibitemOpen
  \bibfield  {author} {\bibinfo {author} {\bibfnamefont {A.~G.}\ \bibnamefont
  {Gurevich}}\ and\ \bibinfo {author} {\bibfnamefont {G.~A.}\ \bibnamefont
  {Melkov}},\ }\href@noop {} {\emph {\bibinfo {title} {Magnetization
  oscillations and waves}}}\ (\bibinfo  {publisher} {CRC press},\ \bibinfo
  {year} {1996})\BibitemShut {NoStop}%
\bibitem [{\citenamefont {Smit}\ and\ \citenamefont
  {Beljers}(1955)}]{Smit:1955}%
  \BibitemOpen
  \bibfield  {author} {\bibinfo {author} {\bibfnamefont {J.}~\bibnamefont
  {Smit}}\ and\ \bibinfo {author} {\bibfnamefont {H.~G.}\ \bibnamefont
  {Beljers}},\ }\bibfield  {title} {\enquote {\bibinfo {title} {Ferromagnetic
  resonance absorption in {B}a{F}e$_{12}${O}$_{19}$ a highly anisotropic
  crystal},}\ }\href@noop {} {\bibfield  {journal} {\bibinfo  {journal}
  {Philips Res. Repts}\ }\textbf {\bibinfo {volume} {10}} (\bibinfo {year}
  {1955})}\BibitemShut {NoStop}%
\bibitem [{\citenamefont {Suhl}(1955)}]{Suhl-PRB:1955}%
  \BibitemOpen
  \bibfield  {author} {\bibinfo {author} {\bibfnamefont {H.}~\bibnamefont
  {Suhl}},\ }\bibfield  {title} {\enquote {\bibinfo {title} {Ferromagnetic
  resonance in nickel ferrite between one and two kilomegacycles},}\ }\href
  {\doibase 10.1103/PhysRev.97.555.2} {\bibfield  {journal} {\bibinfo
  {journal} {Phys. Rev.}\ }\textbf {\bibinfo {volume} {97}},\ \bibinfo {pages}
  {555--557} (\bibinfo {year} {1955})}\BibitemShut {NoStop}%
\bibitem [{\citenamefont {Callen}\ and\ \citenamefont
  {Callen}(1960)}]{Callen-JPCS1960}%
  \BibitemOpen
  \bibfield  {author} {\bibinfo {author} {\bibfnamefont {E.~R.}\ \bibnamefont
  {Callen}}\ and\ \bibinfo {author} {\bibfnamefont {H.~B.}\ \bibnamefont
  {Callen}},\ }\bibfield  {title} {\enquote {\bibinfo {title} {Anisotropic
  magnetization},}\ }\href {\doibase
  https://doi.org/10.1016/0022-3697(60)90161-X} {\bibfield  {journal} {\bibinfo
   {journal} {J. Phys. Chem. Solids}\ }\textbf {\bibinfo {volume} {16}},\
  \bibinfo {pages} {310 -- 328} (\bibinfo {year} {1960})}\BibitemShut {NoStop}%
\bibitem [{\citenamefont {Kittel}\ and\ \citenamefont
  {Van~Vleck}(1960)}]{Kittel:1960}%
  \BibitemOpen
  \bibfield  {author} {\bibinfo {author} {\bibfnamefont {C.}~\bibnamefont
  {Kittel}}\ and\ \bibinfo {author} {\bibfnamefont {J.~H.}\ \bibnamefont
  {Van~Vleck}},\ }\bibfield  {title} {\enquote {\bibinfo {title} {Theory of the
  temperature dependence of the magnetoelastic constants of cubic crystals},}\
  }\href {\doibase 10.1103/PhysRev.118.1231} {\bibfield  {journal} {\bibinfo
  {journal} {Phys. Rev.}\ }\textbf {\bibinfo {volume} {118}},\ \bibinfo {pages}
  {1231--1232} (\bibinfo {year} {1960})}\BibitemShut {NoStop}%
\bibitem [{\citenamefont {Callen}\ and\ \citenamefont
  {Callen}(1965)}]{Callen:PRB1965}%
  \BibitemOpen
  \bibfield  {author} {\bibinfo {author} {\bibfnamefont {E.}~\bibnamefont
  {Callen}}\ and\ \bibinfo {author} {\bibfnamefont {H.~B.}\ \bibnamefont
  {Callen}},\ }\bibfield  {title} {\enquote {\bibinfo {title}
  {Magnetostriction, forced magnetostriction, and anomalous thermal expansion
  in ferromagnets},}\ }\href {\doibase 10.1103/PhysRev.139.A455} {\bibfield
  {journal} {\bibinfo  {journal} {Phys. Rev.}\ }\textbf {\bibinfo {volume}
  {139}},\ \bibinfo {pages} {A455--A471} (\bibinfo {year} {1965})}\BibitemShut
  {NoStop}%
\bibitem [{\citenamefont {O$'$Handley}(1977)}]{OHandley:SSComm1977}%
  \BibitemOpen
  \bibfield  {author} {\bibinfo {author} {\bibfnamefont {R.~C.}\ \bibnamefont
  {O$'$Handley}},\ }\bibfield  {title} {\enquote {\bibinfo {title} {Temperature
  dependence of magnetostriction in {F}e$_{80}${B}$_{20}$ glass},}\ }\href
  {\doibase https://doi.org/10.1016/0038-1098(77)91397-7} {\bibfield  {journal}
  {\bibinfo  {journal} {Solid State Commun.}\ }\textbf {\bibinfo {volume}
  {22}},\ \bibinfo {pages} {485 -- 488} (\bibinfo {year} {1977})}\BibitemShut
  {NoStop}%
\bibitem [{\citenamefont {Barandiar{\'a}n}\ \emph {et~al.}(2011)\citenamefont
  {Barandiar{\'a}n}, \citenamefont {Guti{\'e}rrez},\ and\ \citenamefont
  {Garc{\'i}a-Arribas}}]{Barandiaran:PSS2011}%
  \BibitemOpen
  \bibfield  {author} {\bibinfo {author} {\bibfnamefont {J.~M.}\ \bibnamefont
  {Barandiar{\'a}n}}, \bibinfo {author} {\bibfnamefont {J.}~\bibnamefont
  {Guti{\'e}rrez}}, \ and\ \bibinfo {author} {\bibfnamefont {A.}~\bibnamefont
  {Garc{\'i}a-Arribas}},\ }\bibfield  {title} {\enquote {\bibinfo {title}
  {Magneto-elasticity in amorphous ferromagnets: Basic principles and
  applications},}\ }\href {\doibase https://doi.org/10.1002/pssa.201000738}
  {\bibfield  {journal} {\bibinfo  {journal} {phys. status solidi (a)}\
  }\textbf {\bibinfo {volume} {208}},\ \bibinfo {pages} {2258--2264} (\bibinfo
  {year} {2011})}\BibitemShut {NoStop}%
\bibitem [{\citenamefont {Isogami}\ and\ \citenamefont
  {Taniyama}(2018)}]{Isogami-pps2018}%
  \BibitemOpen
  \bibfield  {author} {\bibinfo {author} {\bibfnamefont {S.}~\bibnamefont
  {Isogami}}\ and\ \bibinfo {author} {\bibfnamefont {T.}~\bibnamefont
  {Taniyama}},\ }\bibfield  {title} {\enquote {\bibinfo {title} {Strain
  mediated in-plane uniaxial magnetic anisotropy in amorphous {C}o{F}e{B} films
  based on structural phase transitions of {B}a{T}i{O}$_3$ single-crystal
  substrates},}\ }\href {\doibase 10.1002/pssa.201700762} {\bibfield  {journal}
  {\bibinfo  {journal} {phys. status solidi (a)}\ }\textbf {\bibinfo {volume}
  {215}},\ \bibinfo {pages} {1700762} (\bibinfo {year} {2018})}\BibitemShut
  {NoStop}%
\bibitem [{\citenamefont {Kats}\ \emph {et~al.}(2016)\citenamefont {Kats},
  \citenamefont {Linnik}, \citenamefont {Salasyuk}, \citenamefont {Rushforth},
  \citenamefont {Wang}, \citenamefont {Wadley}, \citenamefont {Akimov},
  \citenamefont {Cavill}, \citenamefont {Holy}, \citenamefont {Kalashnikova},\
  and\ \citenamefont {Scherbakov}}]{Kats:PRB2016}%
  \BibitemOpen
  \bibfield  {author} {\bibinfo {author} {\bibfnamefont {V.~N.}\ \bibnamefont
  {Kats}}, \bibinfo {author} {\bibfnamefont {T.~L.}\ \bibnamefont {Linnik}},
  \bibinfo {author} {\bibfnamefont {A.~S.}\ \bibnamefont {Salasyuk}}, \bibinfo
  {author} {\bibfnamefont {A.~W.}\ \bibnamefont {Rushforth}}, \bibinfo {author}
  {\bibfnamefont {M.}~\bibnamefont {Wang}}, \bibinfo {author} {\bibfnamefont
  {P.}~\bibnamefont {Wadley}}, \bibinfo {author} {\bibfnamefont {A.~V.}\
  \bibnamefont {Akimov}}, \bibinfo {author} {\bibfnamefont {S.~A.}\
  \bibnamefont {Cavill}}, \bibinfo {author} {\bibfnamefont {V.}~\bibnamefont
  {Holy}}, \bibinfo {author} {\bibfnamefont {A.~M.}\ \bibnamefont
  {Kalashnikova}}, \ and\ \bibinfo {author} {\bibfnamefont {A.~V.}\
  \bibnamefont {Scherbakov}},\ }\bibfield  {title} {\enquote {\bibinfo {title}
  {Ultrafast changes of magnetic anisotropy driven by laser-generated coherent
  and noncoherent phonons in metallic films},}\ }\href {\doibase
  10.1103/PhysRevB.93.214422} {\bibfield  {journal} {\bibinfo  {journal} {Phys.
  Rev. B}\ }\textbf {\bibinfo {volume} {93}},\ \bibinfo {pages} {214422}
  (\bibinfo {year} {2016})}\BibitemShut {NoStop}%
\bibitem [{\citenamefont {J\"ager}\ \emph {et~al.}(2015)\citenamefont
  {J\"ager}, \citenamefont {Scherbakov}, \citenamefont {Glavin}, \citenamefont
  {Salasyuk}, \citenamefont {Campion}, \citenamefont {Rushforth}, \citenamefont
  {Yakovlev}, \citenamefont {Akimov},\ and\ \citenamefont
  {Bayer}}]{Jager:PRB2015}%
  \BibitemOpen
  \bibfield  {author} {\bibinfo {author} {\bibfnamefont {J.~V.}\ \bibnamefont
  {J\"ager}}, \bibinfo {author} {\bibfnamefont {A.~V.}\ \bibnamefont
  {Scherbakov}}, \bibinfo {author} {\bibfnamefont {B.~A.}\ \bibnamefont
  {Glavin}}, \bibinfo {author} {\bibfnamefont {A.~S.}\ \bibnamefont
  {Salasyuk}}, \bibinfo {author} {\bibfnamefont {R.~P.}\ \bibnamefont
  {Campion}}, \bibinfo {author} {\bibfnamefont {A.~W.}\ \bibnamefont
  {Rushforth}}, \bibinfo {author} {\bibfnamefont {D.~R.}\ \bibnamefont
  {Yakovlev}}, \bibinfo {author} {\bibfnamefont {A.~V.}\ \bibnamefont
  {Akimov}}, \ and\ \bibinfo {author} {\bibfnamefont {M.}~\bibnamefont
  {Bayer}},\ }\bibfield  {title} {\enquote {\bibinfo {title}
  {\textcolor{newtext}{Resonant driving of magnetization precession in a
  ferromagnetic layer by coherent monochromatic phonons}},}\ }\href {\doibase
  10.1103/PhysRevB.92.020404} {\bibfield  {journal} {\bibinfo  {journal} {Phys.
  Rev. B}\ }\textbf {\bibinfo {volume} {92}},\ \bibinfo {pages} {020404(R)}
  (\bibinfo {year} {2015})}\BibitemShut {NoStop}%
\bibitem [{\citenamefont {Carpene}\ \emph {et~al.}(2011)\citenamefont
  {Carpene}, \citenamefont {Piovera}, \citenamefont {Dallera}, \citenamefont
  {Mancini},\ and\ \citenamefont {Puppin}}]{Carpene:PRB2011}%
  \BibitemOpen
  \bibfield  {author} {\bibinfo {author} {\bibfnamefont {E.}~\bibnamefont
  {Carpene}}, \bibinfo {author} {\bibfnamefont {C.}~\bibnamefont {Piovera}},
  \bibinfo {author} {\bibfnamefont {C.}~\bibnamefont {Dallera}}, \bibinfo
  {author} {\bibfnamefont {E.}~\bibnamefont {Mancini}}, \ and\ \bibinfo
  {author} {\bibfnamefont {E.}~\bibnamefont {Puppin}},\ }\bibfield  {title}
  {\enquote {\bibinfo {title} {\textcolor{newtext}{All-optical subnanosecond
  coherent spin switching in thin ferromagnetic layers}},}\ }\href {\doibase
  10.1103/PhysRevB.84.134425} {\bibfield  {journal} {\bibinfo  {journal} {Phys.
  Rev. B}\ }\textbf {\bibinfo {volume} {84}},\ \bibinfo {pages} {134425}
  (\bibinfo {year} {2011})}\BibitemShut {NoStop}%
\bibitem [{\citenamefont {Stupakiewicz}\ \emph {et~al.}(2017)\citenamefont
  {Stupakiewicz}, \citenamefont {Szerenos}, \citenamefont {Afanasiev},
  \citenamefont {Kirilyuk},\ and\ \citenamefont
  {Kimel}}]{Stupakiewicz:Nature2017}%
  \BibitemOpen
  \bibfield  {author} {\bibinfo {author} {\bibfnamefont {A.}~\bibnamefont
  {Stupakiewicz}}, \bibinfo {author} {\bibfnamefont {K.}~\bibnamefont
  {Szerenos}}, \bibinfo {author} {\bibfnamefont {D.}~\bibnamefont {Afanasiev}},
  \bibinfo {author} {\bibfnamefont {A.}~\bibnamefont {Kirilyuk}}, \ and\
  \bibinfo {author} {\bibfnamefont {A.~V.}\ \bibnamefont {Kimel}},\ }\bibfield
  {title} {\enquote {\bibinfo {title} {\textcolor{newtext}{Ultrafast nonthermal
  photo-magnetic recording in a transparent medium}},}\ }\href {\doibase
  10.1038/nature20807} {\bibfield  {journal} {\bibinfo  {journal} {Nature}\
  }\textbf {\bibinfo {volume} {542}},\ \bibinfo {pages} {71} (\bibinfo {year}
  {2017})}\BibitemShut {NoStop}%
\bibitem [{\citenamefont {Davies}\ \emph {et~al.}(2019)\citenamefont {Davies},
  \citenamefont {Prabhakara}, \citenamefont {Davydova}, \citenamefont
  {Zvezdin}, \citenamefont {Shapaeva}, \citenamefont {Wang}, \citenamefont
  {Zvezdin}, \citenamefont {Kirilyuk}, \citenamefont {Rasing},\ and\
  \citenamefont {Kimel}}]{Davies:PRL2019}%
  \BibitemOpen
  \bibfield  {author} {\bibinfo {author} {\bibfnamefont {C.~S.}\ \bibnamefont
  {Davies}}, \bibinfo {author} {\bibfnamefont {K.~H.}\ \bibnamefont
  {Prabhakara}}, \bibinfo {author} {\bibfnamefont {M.~D.}\ \bibnamefont
  {Davydova}}, \bibinfo {author} {\bibfnamefont {K.~A.}\ \bibnamefont
  {Zvezdin}}, \bibinfo {author} {\bibfnamefont {T.~B.}\ \bibnamefont
  {Shapaeva}}, \bibinfo {author} {\bibfnamefont {S.}~\bibnamefont {Wang}},
  \bibinfo {author} {\bibfnamefont {A.~K.}\ \bibnamefont {Zvezdin}}, \bibinfo
  {author} {\bibfnamefont {A.}~\bibnamefont {Kirilyuk}}, \bibinfo {author}
  {\bibfnamefont {Th.}\ \bibnamefont {Rasing}}, \ and\ \bibinfo {author}
  {\bibfnamefont {A.~V.}\ \bibnamefont {Kimel}},\ }\bibfield  {title} {\enquote
  {\bibinfo {title} {\textcolor{newtext}{Anomalously Damped Heat-Assisted Route
  for Precessional Magnetization Reversal in an Iron Garnet}},}\ }\href
  {\doibase 10.1103/PhysRevLett.122.027202} {\bibfield  {journal} {\bibinfo
  {journal} {Phys. Rev. Lett.}\ }\textbf {\bibinfo {volume} {122}},\ \bibinfo
  {pages} {027202} (\bibinfo {year} {2019})}\BibitemShut {NoStop}%
\bibitem [{\citenamefont {Tudosa}\ \emph {et~al.}(2004)\citenamefont {Tudosa},
  \citenamefont {Stamm}, \citenamefont {Kashuba}, \citenamefont {King},
  \citenamefont {Siegmann}, \citenamefont {Stöhr}, \citenamefont {Ju},
  \citenamefont {Lu},\ and\ \citenamefont {Weller}}]{Tudosa:Nature2004}%
  \BibitemOpen
  \bibfield  {author} {\bibinfo {author} {\bibfnamefont {I.}~\bibnamefont
  {Tudosa}}, \bibinfo {author} {\bibfnamefont {C.}~\bibnamefont {Stamm}},
  \bibinfo {author} {\bibfnamefont {A.~B.}\ \bibnamefont {Kashuba}}, \bibinfo
  {author} {\bibfnamefont {F.}~\bibnamefont {King}}, \bibinfo {author}
  {\bibfnamefont {H.~C.}\ \bibnamefont {Siegmann}}, \bibinfo {author}
  {\bibfnamefont {J.}~\bibnamefont {Stöhr}}, \bibinfo {author} {\bibfnamefont
  {G.}~\bibnamefont {Ju}}, \bibinfo {author} {\bibfnamefont {B.}~\bibnamefont
  {Lu}}, \ and\ \bibinfo {author} {\bibfnamefont {D.}~\bibnamefont {Weller}},\
  }\bibfield  {title} {\enquote {\bibinfo {title} {\textcolor{newtext}{The
  ultimate speed of magnetic switching in granular recording media}},}\ }\href
  {\doibase 10.1038/nature02438} {\bibfield  {journal} {\bibinfo  {journal}
  {Nature}\ }\textbf {\bibinfo {volume} {428}},\ \bibinfo {pages} {831}
  (\bibinfo {year} {2004})}\BibitemShut {NoStop}%
\bibitem [{\citenamefont {de~Jong}\ \emph {et~al.}(2012)\citenamefont
  {de~Jong}, \citenamefont {Razdolski}, \citenamefont {Kalashnikova},
  \citenamefont {Pisarev}, \citenamefont {Balbashov}, \citenamefont {Kirilyuk},
  \citenamefont {Rasing},\ and\ \citenamefont {Kimel}}]{deJong:PRL2012}%
  \BibitemOpen
  \bibfield  {author} {\bibinfo {author} {\bibfnamefont {J.~A.}\ \bibnamefont
  {de~Jong}}, \bibinfo {author} {\bibfnamefont {I.}~\bibnamefont {Razdolski}},
  \bibinfo {author} {\bibfnamefont {A.~M.}\ \bibnamefont {Kalashnikova}},
  \bibinfo {author} {\bibfnamefont {R.~V.}\ \bibnamefont {Pisarev}}, \bibinfo
  {author} {\bibfnamefont {A.~M.}\ \bibnamefont {Balbashov}}, \bibinfo {author}
  {\bibfnamefont {A.}~\bibnamefont {Kirilyuk}}, \bibinfo {author}
  {\bibfnamefont {Th.}\ \bibnamefont {Rasing}}, \ and\ \bibinfo {author}
  {\bibfnamefont {A.~V.}\ \bibnamefont {Kimel}},\ }\bibfield  {title} {\enquote
  {\bibinfo {title} {\textcolor{newtext}{Coherent Control of the Route of an
  Ultrafast Magnetic Phase Transition via Low-Amplitude Spin Precession}},}\
  }\href {\doibase 10.1103/PhysRevLett.108.157601} {\bibfield  {journal}
  {\bibinfo  {journal} {Phys. Rev. Lett.}\ }\textbf {\bibinfo {volume} {108}},\
  \bibinfo {pages} {157601} (\bibinfo {year} {2012})}\BibitemShut {NoStop}%
\bibitem [{\citenamefont {Zhang}\ \emph {et~al.}(2011)\citenamefont {Zhang},
  \citenamefont {Fan}, \citenamefont {Wang}, \citenamefont {Kou}, \citenamefont
  {Cao}, \citenamefont {Chen}, \citenamefont {Ni}, \citenamefont {Pan},\ and\
  \citenamefont {Xiao}}]{Zhang:APL2011}%
  \BibitemOpen
  \bibfield  {author} {\bibinfo {author} {\bibfnamefont {Y.}~\bibnamefont
  {Zhang}}, \bibinfo {author} {\bibfnamefont {X.}~\bibnamefont {Fan}}, \bibinfo
  {author} {\bibfnamefont {W.}~\bibnamefont {Wang}}, \bibinfo {author}
  {\bibfnamefont {X.}~\bibnamefont {Kou}}, \bibinfo {author} {\bibfnamefont
  {R.}~\bibnamefont {Cao}}, \bibinfo {author} {\bibfnamefont {X.}~\bibnamefont
  {Chen}}, \bibinfo {author} {\bibfnamefont {Ch.}\ \bibnamefont {Ni}}, \bibinfo
  {author} {\bibfnamefont {L.}~\bibnamefont {Pan}}, \ and\ \bibinfo {author}
  {\bibfnamefont {J.~Q.}\ \bibnamefont {Xiao}},\ }\bibfield  {title} {\enquote
  {\bibinfo {title} {\textcolor{newtext}{Study and tailoring spin dynamic
  properties of {C}o{F}e{B} during rapid thermal annealing}},}\ }\href
  {\doibase 10.1063/1.3549188} {\bibfield  {journal} {\bibinfo  {journal}
  {Appl. Phys. Lett.}\ }\textbf {\bibinfo {volume} {98}},\ \bibinfo {pages}
  {042506} (\bibinfo {year} {2011})}\BibitemShut {NoStop}%
\bibitem [{\citenamefont {Srivastava}\ \emph {et~al.}(2018)\citenamefont
  {Srivastava}, \citenamefont {Chen}, \citenamefont {Dutta}, \citenamefont
  {Ramaswamy}, \citenamefont {Son}, \citenamefont {Saifullah}, \citenamefont
  {Yamane}, \citenamefont {Lee}, \citenamefont {Teo}, \citenamefont {Feng},\
  and\ \citenamefont {Yang}}]{Srivastava:PRApp2018}%
  \BibitemOpen
  \bibfield  {author} {\bibinfo {author} {\bibfnamefont {S.}~\bibnamefont
  {Srivastava}}, \bibinfo {author} {\bibfnamefont {A.~P.}\ \bibnamefont
  {Chen}}, \bibinfo {author} {\bibfnamefont {T.}~\bibnamefont {Dutta}},
  \bibinfo {author} {\bibfnamefont {R.}~\bibnamefont {Ramaswamy}}, \bibinfo
  {author} {\bibfnamefont {J.}~\bibnamefont {Son}}, \bibinfo {author}
  {\bibfnamefont {M.~S.~M.}\ \bibnamefont {Saifullah}}, \bibinfo {author}
  {\bibfnamefont {K.}~\bibnamefont {Yamane}}, \bibinfo {author} {\bibfnamefont
  {K.}~\bibnamefont {Lee}}, \bibinfo {author} {\bibfnamefont {K.-L.}\
  \bibnamefont {Teo}}, \bibinfo {author} {\bibfnamefont {Y.~P.}\ \bibnamefont
  {Feng}}, \ and\ \bibinfo {author} {\bibfnamefont {H.}~\bibnamefont {Yang}},\
  }\bibfield  {title} {\enquote {\bibinfo {title} {\textcolor{newtext}{ Effect
  of
  $({\mathrm{Co}}_{x}{\mathrm{Fe}}_{1\ensuremath{-}x}{)}_{80}{\mathrm{B}}_{20}$
  Composition on the Magnetic Properties of the Free Layer in Double-Barrier
  Magnetic Tunnel Junctions}},}\ }\href {\doibase
  10.1103/PhysRevApplied.10.024031} {\bibfield  {journal} {\bibinfo  {journal}
  {Phys. Rev. Applied}\ }\textbf {\bibinfo {volume} {10}},\ \bibinfo {pages}
  {024031} (\bibinfo {year} {2018})}\BibitemShut {NoStop}%
\bibitem [{\citenamefont {Khokhlov}\ \emph {et~al.}(2019)\citenamefont
  {Khokhlov}, \citenamefont {Gerevenkov}, \citenamefont {Shelukhin},
  \citenamefont {Azovtsev}, \citenamefont {Pertsev}, \citenamefont {Wang},
  \citenamefont {Rushforth}, \citenamefont {Scherbakov},\ and\ \citenamefont
  {Kalashnikova}}]{Khokhlov:2018}%
  \BibitemOpen
  \bibfield  {author} {\bibinfo {author} {\bibfnamefont {N.~E.}\ \bibnamefont
  {Khokhlov}}, \bibinfo {author} {\bibfnamefont {P.~I.}\ \bibnamefont
  {Gerevenkov}}, \bibinfo {author} {\bibfnamefont {L.~A.}\ \bibnamefont
  {Shelukhin}}, \bibinfo {author} {\bibfnamefont {A.~V.}\ \bibnamefont
  {Azovtsev}}, \bibinfo {author} {\bibfnamefont {N.~A.}\ \bibnamefont
  {Pertsev}}, \bibinfo {author} {\bibfnamefont {M.}~\bibnamefont {Wang}},
  \bibinfo {author} {\bibfnamefont {A.~W.}\ \bibnamefont {Rushforth}}, \bibinfo
  {author} {\bibfnamefont {A.~V.}\ \bibnamefont {Scherbakov}}, \ and\ \bibinfo
  {author} {\bibfnamefont {A.~M.}\ \bibnamefont {Kalashnikova}},\ }\bibfield
  {title} {\enquote {\bibinfo {title} {Optical excitation of propagating
  magnetostatic waves in an epitaxial galfenol film by ultrafast magnetic
  anisotropy change},}\ }\href {\doibase 10.1103/PhysRevApplied.12.044044}
  {\bibfield  {journal} {\bibinfo  {journal} {Phys. Rev. Appl.}\ }\textbf
  {\bibinfo {volume} {12}},\ \bibinfo {pages} {044044} (\bibinfo {year}
  {2019})}\BibitemShut {NoStop}%
\bibitem [{\citenamefont {Van~de Wiele}\ \emph {et~al.}(2016)\citenamefont
  {Van~de Wiele}, \citenamefont {H{\"a}m{\"a}l{\"a}inen}, \citenamefont
  {Bal{\'a}{\v{z}}}, \citenamefont {Montoncello},\ and\ \citenamefont
  {Van~Dijken}}]{VanDeWiele:2016}%
  \BibitemOpen
  \bibfield  {author} {\bibinfo {author} {\bibfnamefont {B.}~\bibnamefont
  {Van~de Wiele}}, \bibinfo {author} {\bibfnamefont {S.~J.}\ \bibnamefont
  {H{\"a}m{\"a}l{\"a}inen}}, \bibinfo {author} {\bibfnamefont {P.}~\bibnamefont
  {Bal{\'a}{\v{z}}}}, \bibinfo {author} {\bibfnamefont {F.}~\bibnamefont
  {Montoncello}}, \ and\ \bibinfo {author} {\bibfnamefont {S.}~\bibnamefont
  {Van~Dijken}},\ }\bibfield  {title} {\enquote {\bibinfo {title} {Tunable
  short-wavelength spin wave excitation from pinned magnetic domain walls},}\
  }\href {\doibase https://doi.org/10.1038/srep21330} {\bibfield  {journal}
  {\bibinfo  {journal} {Sci. Rep.}\ }\textbf {\bibinfo {volume} {6}},\ \bibinfo
  {pages} {21330} (\bibinfo {year} {2016})}\BibitemShut {NoStop}%
\bibitem [{\citenamefont {L{\'o}pez~Gonz{\'a}lez}\ \emph
  {et~al.}(2016)\citenamefont {L{\'o}pez~Gonz{\'a}lez}, \citenamefont
  {Casiraghi}, \citenamefont {Van~de Wiele},\ and\ \citenamefont
  {Van~Dijken}}]{LopezGonzalez:2016}%
  \BibitemOpen
  \bibfield  {author} {\bibinfo {author} {\bibfnamefont {D.}~\bibnamefont
  {L{\'o}pez~Gonz{\'a}lez}}, \bibinfo {author} {\bibfnamefont {A.}~\bibnamefont
  {Casiraghi}}, \bibinfo {author} {\bibfnamefont {B.}~\bibnamefont {Van~de
  Wiele}}, \ and\ \bibinfo {author} {\bibfnamefont {S.}~\bibnamefont
  {Van~Dijken}},\ }\bibfield  {title} {\enquote {\bibinfo {title}
  {Reconfigurable magnetic logic based on the energetics of pinned domain
  walls},}\ }\href {\doibase https://doi.org/10.1063/1.4940119} {\bibfield
  {journal} {\bibinfo  {journal} {Appl. Phys. Lett.}\ }\textbf {\bibinfo
  {volume} {108}},\ \bibinfo {pages} {032402} (\bibinfo {year}
  {2016})}\BibitemShut {NoStop}%
\bibitem [{\citenamefont {Zvezdin}\ and\ \citenamefont
  {Kotov}(1997)}]{Zvezdin_book}%
  \BibitemOpen
  \bibfield  {author} {\bibinfo {author} {\bibfnamefont {A.~K.}\ \bibnamefont
  {Zvezdin}}\ and\ \bibinfo {author} {\bibfnamefont {V.~K.}\ \bibnamefont
  {Kotov}},\ }\href@noop {} {\emph {\bibinfo {title}
  {\textcolor{newtext}{Modern Magnetooptics and Magnetooptical Materials}}}}\
  (\bibinfo  {publisher} {IoP Publishing},\ \bibinfo {year} {1997})\BibitemShut
  {NoStop}%
\bibitem [{\citenamefont {Liang}\ \emph {et~al.}(2015)\citenamefont {Liang},
  \citenamefont {Xu}, \citenamefont {Zheng}, \citenamefont {Lum},\ and\
  \citenamefont {Qiu}}]{Liang:OSA2015}%
  \BibitemOpen
  \bibfield  {author} {\bibinfo {author} {\bibfnamefont {X.}~\bibnamefont
  {Liang}}, \bibinfo {author} {\bibfnamefont {X.}~\bibnamefont {Xu}}, \bibinfo
  {author} {\bibfnamefont {R.}~\bibnamefont {Zheng}}, \bibinfo {author}
  {\bibfnamefont {Z.~A.}\ \bibnamefont {Lum}}, \ and\ \bibinfo {author}
  {\bibfnamefont {J.}~\bibnamefont {Qiu}},\ }\bibfield  {title} {\enquote
  {\bibinfo {title} {Optical constant of {C}o{F}e{B} thin film measured with
  the interference enhancement method},}\ }\href {\doibase
  10.1364/AO.54.001557} {\bibfield  {journal} {\bibinfo  {journal} {Appl.
  Opt.}\ }\textbf {\bibinfo {volume} {54}},\ \bibinfo {pages} {1557--1563}
  (\bibinfo {year} {2015})}\BibitemShut {NoStop}%
\bibitem [{\citenamefont {Hoffmann}\ \emph {et~al.}(2019)\citenamefont
  {Hoffmann}, \citenamefont {Sharma}, \citenamefont {Matthes}, \citenamefont
  {Okano}, \citenamefont {Hellwig}, \citenamefont {Ecke}, \citenamefont {Zahn},
  \citenamefont {Salvan},\ and\ \citenamefont {Schulz}}]{Hoffmann:2019}%
  \BibitemOpen
  \bibfield  {author} {\bibinfo {author} {\bibfnamefont {M.~A.}\ \bibnamefont
  {Hoffmann}}, \bibinfo {author} {\bibfnamefont {A.}~\bibnamefont {Sharma}},
  \bibinfo {author} {\bibfnamefont {P.}~\bibnamefont {Matthes}}, \bibinfo
  {author} {\bibfnamefont {S.}~\bibnamefont {Okano}}, \bibinfo {author}
  {\bibfnamefont {O.}~\bibnamefont {Hellwig}}, \bibinfo {author} {\bibfnamefont
  {R}~\bibnamefont {Ecke}}, \bibinfo {author} {\bibfnamefont {D.~R.~T.}\
  \bibnamefont {Zahn}}, \bibinfo {author} {\bibfnamefont {G.}~\bibnamefont
  {Salvan}}, \ and\ \bibinfo {author} {\bibfnamefont {S.~E.}\ \bibnamefont
  {Schulz}},\ }\bibfield  {title} {\enquote {\bibinfo {title}
  {\textcolor{newtext}{Spectroscopic ellipsometry and magneto-optical Kerr
  effect spectroscopy study of thermally treated Co$_{60}$Fe$_{20}$B$_{20}$
  thin films}},}\ }\href {\doibase 10.1088/1361-648x/ab4d2f} {\bibfield
  {journal} {\bibinfo  {journal} {J. Phys. Condens. Matter}\ }\textbf {\bibinfo
  {volume} {32}},\ \bibinfo {pages} {055702} (\bibinfo {year}
  {2019})}\BibitemShut {NoStop}%
\bibitem [{\citenamefont {Rosenblatt}\ \emph {et~al.}(2020)\citenamefont
  {Rosenblatt}, \citenamefont {Simkhovich}, \citenamefont {Bartal},\ and\
  \citenamefont {Orenstein}}]{Rosenblatt:PRX2020}%
  \BibitemOpen
  \bibfield  {author} {\bibinfo {author} {\bibfnamefont {G.}~\bibnamefont
  {Rosenblatt}}, \bibinfo {author} {\bibfnamefont {B.}~\bibnamefont
  {Simkhovich}}, \bibinfo {author} {\bibfnamefont {G.}~\bibnamefont {Bartal}},
  \ and\ \bibinfo {author} {\bibfnamefont {M.}~\bibnamefont {Orenstein}},\
  }\bibfield  {title} {\enquote {\bibinfo {title} {\textcolor{newtext}{Nonmodal
  Plasmonics: Controlling the Forced Optical Response of Nanostructures}},}\
  }\href {\doibase 10.1103/PhysRevX.10.011071} {\bibfield  {journal} {\bibinfo
  {journal} {Phys. Rev. X}\ }\textbf {\bibinfo {volume} {10}},\ \bibinfo
  {pages} {011071} (\bibinfo {year} {2020})}\BibitemShut {NoStop}%
\bibitem [{\citenamefont {Walter}\ \emph {et~al.}(2011)\citenamefont {Walter},
  \citenamefont {Walowski}, \citenamefont {Zbarsky}, \citenamefont
  {M{\"u}nzenberg}, \citenamefont {Sch{\"a}fers}, \citenamefont {Ebke},
  \citenamefont {Reiss}, \citenamefont {Thomas}, \citenamefont {Peretzki},
  \citenamefont {Seibt}, \citenamefont {Moodera}, \citenamefont {Czerner},
  \citenamefont {Bachmann},\ and\ \citenamefont
  {Heiliger}}]{walter:natMat2011}%
  \BibitemOpen
  \bibfield  {author} {\bibinfo {author} {\bibfnamefont {M.}~\bibnamefont
  {Walter}}, \bibinfo {author} {\bibfnamefont {J.}~\bibnamefont {Walowski}},
  \bibinfo {author} {\bibfnamefont {V.}~\bibnamefont {Zbarsky}}, \bibinfo
  {author} {\bibfnamefont {M.}~\bibnamefont {M{\"u}nzenberg}}, \bibinfo
  {author} {\bibfnamefont {M.}~\bibnamefont {Sch{\"a}fers}}, \bibinfo {author}
  {\bibfnamefont {D.}~\bibnamefont {Ebke}}, \bibinfo {author} {\bibfnamefont
  {G.}~\bibnamefont {Reiss}}, \bibinfo {author} {\bibfnamefont
  {A.}~\bibnamefont {Thomas}}, \bibinfo {author} {\bibfnamefont
  {P.}~\bibnamefont {Peretzki}}, \bibinfo {author} {\bibfnamefont
  {M.}~\bibnamefont {Seibt}}, \bibinfo {author} {\bibfnamefont {J.~S.}\
  \bibnamefont {Moodera}}, \bibinfo {author} {\bibfnamefont {M.}~\bibnamefont
  {Czerner}}, \bibinfo {author} {\bibfnamefont {M.}~\bibnamefont {Bachmann}}, \
  and\ \bibinfo {author} {\bibfnamefont {C.}~\bibnamefont {Heiliger}},\
  }\bibfield  {title} {\enquote {\bibinfo {title} {Seebeck effect in magnetic
  tunnel junctions},}\ }\href {\doibase https://doi.org/10.1038/nmat3076}
  {\bibfield  {journal} {\bibinfo  {journal} {Nat. Mater.}\ }\textbf {\bibinfo
  {volume} {10}},\ \bibinfo {pages} {742} (\bibinfo {year} {2011})}\BibitemShut
  {NoStop}%
\bibitem [{\citenamefont {O$'$Handley}\ \emph {et~al.}(1976)\citenamefont
  {O$'$Handley}, \citenamefont {Hasegawa}, \citenamefont {Ray},\ and\
  \citenamefont {Chou}}]{oHandley:APL1976}%
  \BibitemOpen
  \bibfield  {author} {\bibinfo {author} {\bibfnamefont {R.~C.}\ \bibnamefont
  {O$'$Handley}}, \bibinfo {author} {\bibfnamefont {R.}~\bibnamefont
  {Hasegawa}}, \bibinfo {author} {\bibfnamefont {R.}~\bibnamefont {Ray}}, \
  and\ \bibinfo {author} {\bibfnamefont {C.-P.}\ \bibnamefont {Chou}},\
  }\bibfield  {title} {\enquote {\bibinfo {title} {Ferromagnetic properties of
  some new metallic glasses},}\ }\href {\doibase
  https://doi.org/10.1063/1.89085} {\bibfield  {journal} {\bibinfo  {journal}
  {Appl. Phys. Lett.}\ }\textbf {\bibinfo {volume} {29}},\ \bibinfo {pages}
  {330--332} (\bibinfo {year} {1976})}\BibitemShut {NoStop}%
\bibitem [{\citenamefont {Walowski}(2012)}]{walowski2012PhD}%
  \BibitemOpen
  \bibfield  {author} {\bibinfo {author} {\bibfnamefont {J.}~\bibnamefont
  {Walowski}},\ }\emph {\bibinfo {title} {\textcolor{newtext}{Physics of laser
  heated ferromagnets: Ultrafast demagnetization and magneto-{S}eebeck
  effect}}},\ \href@noop {} {Ph.D. thesis},\ \bibinfo  {school}
  {Nieders{\"a}chsische Staats-und Universit{\"a}tsbibliothek G{\"o}ttingen}
  (\bibinfo {year} {2012})\BibitemShut {NoStop}%
\end{thebibliography}%

\end{document}